\documentclass[aps,physrev, twocolumn, superscriptaddress]{revtex4-2}
\usepackage{graphicx}
\usepackage{longtable}
\usepackage{amsmath}
\usepackage[overload]{textcase} 

\begin{document}

\title{Structural phase transitions in the van der Waals ferromagnets Fe$_x$Pd$_{y}$Te$_2$}

\author{R. F. S. Penacchio}
\affiliation{Ames National Laboratory, US DOE, Iowa State University, Ames, IA, USA}
\affiliation{Department of Physics, Iowa State University, Ames, Iowa 50011, USA}
\affiliation{Institute of Physics, University of S{\~{a}}o Paulo, S{\~{a}}o Paulo, SP, Brazil}

\author{S. Mohamed}
\affiliation{Ames National Laboratory, US DOE, Iowa State University, Ames, IA, USA}
\affiliation{Department of Chemistry, Iowa State University, Ames, Iowa 50011, USA}

\author{S{\'{e}}rgio L. Morelh{\~{a}}o}
\affiliation{Institute of Physics, University of S{\~{a}}o Paulo, S{\~{a}}o Paulo, SP, Brazil}

\author{S. L. Bud’ko}
\affiliation{Ames National Laboratory, US DOE, Iowa State University, Ames, IA, USA}
\affiliation{Department of Physics, Iowa State University, Ames, Iowa 50011, USA}

\author{P. C. Canfield}
\affiliation{Ames National Laboratory, US DOE, Iowa State University, Ames, IA, USA}
\affiliation{Department of Physics, Iowa State University, Ames, Iowa 50011, USA}

\author{T. J. Slade}
\affiliation{Ames National Laboratory, US DOE, Iowa State University, Ames, IA, USA}

\date{\today}

\begin{abstract}

Two-dimensional van der Waals (vdW) magnets are attracting significant attention, both as platforms for studying fundamental magnetic interactions and for the exciting possibility of utilizing them as building blocks in devices and heterostructures, which may lead to new physical phenomena and functionalities. Here, we provide a detailed study of the crystal structure and physical properties of the recently discovered vdW ferromagnet FePd$_2$Te$_2$. We find this compound has a relatively wide width of formation, and grow single crystals with compositions Fe$_x$Pd$_{y}$Te$_2$ where $x$ ranges from 0.9 to 1.1 and $y$ from 1.8 to 2.5, respectively. Temperature-dependent X-ray diffraction and transport measurements reveal that a first-order structural transition occurs in the range of $T$ = 360-420 K, where the critical temperature, modulation wave vector, and corresponding room-temperature crystal structures all depend on chemical composition. Above the transition, the compounds with Pd fraction $y>2$ adopt a disordered derivative of the tetragonal FeTe structure, with the Fe layer showing mixed Fe/Pd occupancy and the extra Pd atoms partially occupying interstitial sites. Below 370 K, the structure is incommensurately modulated, likely associated with the complex ordering of Pd/Fe atoms in the metal layers or the interstitial Pd in the vdW gaps. For $y<2$, the composition Fe$_{1.1}$Pd$_{1.8}$Te$_2$ has monoclinic symmetry at room temperature that is consistent with the reported structure of FePd$_2$Te$_2$. This phase undergoes a structural transition at 420 K for which the high temperature structure is yet to be determined; however, based on the similarities with the $y > 2$ compounds, we speculate that this composition also adopts a tetragonal structure above 420\,K.  Importantly, the high temperature, symmetry-breaking structural transition observed here provides a possible explanation for the origin of the structural domains previously observed in FePd$_2$Te$_2$. All compounds investigated in the Fe$_x$Pd$_{y}$Te$_2$ series show metallic behavior, with magnetic characterization indicating that they are easy-plane, hard, ferromagnets with $T_C$ spanning 98--180 K. Both the critical temperature for the structural transition and the Curie temperature are moderately suppressed with increasing Pd fraction $y$ and corresponding decreasing Fe fraction $x$, indicating that synthetic control over $x$ and $y$ paves way for the further exploration of these compounds.

\end{abstract}

\maketitle

\section{Introduction}

Intrinsic magnetism in two-dimensional (2D) van der Waals (vdW) materials has generated intense interest \cite{Park_2016,sachs2013ferromagnetic,lee2016ising,ellner,williams2015magnetic, huang2017layer,bonilla2018strong,gong2017,deiseroth2006,burch2018, seyler2018,ziebel2024crsbr, lee2024magnetic, idzuchi2021unconventional}, as such materials are an ideal platform for studying fundamental magnetic interactions and phase transitions in the 2D limit, where fluctuations are expected to be strongly enhanced \cite{PhysRevLett.17.1133,PhysRev.65.117,PhysRevLett.23.861,posey2024two}. Beyond this, magnetic vdW materials offer tremendous technological potential as components in microelectronic and spintronic devices. This is due to the exciting possibility of stacking these compounds in heterostructures, where layer composition, layer number, and relative orientations represent adjustable parameters that may allow the tuning of properties, resulting in different functionalities, magnetic ordering and/or new physical phenomena \cite {macneill2017control,qian2014quantum,gibertini2019,deng2018, wang2018,song2018giant,xing2017electric}.

FePd$_2$Te$_2$ is a recently discovered vdW ferromagnet that adopts a new structure type with monoclinic symmetry \cite{shi2024}. The structure is an ordered derivative of the FeTe prototype, in which the central Fe layer is composed of alternating Fe-Fe and Pd-Pd zigzag chains, in addition to 8\%  occupancy of Pd at interstitial sites between the vdW gaps. Interestingly, both the report outlining the discovery of this compound and more recent publications show that the FePd$_2$Te$_2$ crystals exhibit heavy twinning at the sub-micron scale \cite{shi2024, chen2024, mi2025,ruiz2025tunable}, resulting in perpendicular intersections between the Fe chains in adjacent domains. Whereas the twinning produces a macroscopic spin texture that may support interesting magnetotransport phenomena \cite{chen2024, li2025anomalous, ruiz2025tunable}, the origin of the structural twins remains an open question.

In this work, we explore the crystal structure and physical properties of FePd$_2$Te$_2$-based single crystals. In contrast to the stoichiometric 122 composition, we find that these materials have a flexible composition Fe$_{x}$Pd$_{y}$Te$_2$, in which the Fe fraction typically decreases with increasing Pd. Here, we investigated samples with $x$ = 0.9--1.1 and $y$ = 1.8--2.5. Based on high temperature X-ray diffraction, resistance, and magnetization measurements, we show that the Fe$_x$Pd$_y$Te$_2$ series undergoes a symmetry-breaking structural phase transition at temperatures between 360-420 K, depending on the specific composition. Above the transition, the $y>2$ compounds adopt a heavily disordered variant of the tetragonal FeTe structure (space group $P$4/$nmm$), with the Fe layer showing nearly 50:50 mixed Fe/Pd occupancy and the remaining Pd accommodated into the interstitial positions between vdW gaps. For $y>2$, the high-temperature structure of Fe$_{1.1}$Pd$_{1.8}$Te$_2$ is yet to be determined, but its low-temperature structure is consistent with that reported for FePd$_2$Te$_2$ \cite{shi2024}. Below the transition, we observe satellite peaks for all compositions and find that the modulation wave vector is also sensitive to the chemical composition. Specifically, we find a commensurate wave vector of \textit{q} = ($\frac{1}{2}$ $\frac{1}{2}$ $\frac{1}{2}$) for Fe$_{1.1}$Pd$_{1.8}$Te$_2$, whereas samples with Pd fraction $y >$ 2 are incommensurately modulated. Our identification of a symmetry-lowering structural transition from a high-temperature tetragonal phase likely explains the origin of the rich structural domains reported in earlier work \cite{shi2024,chen2024, mi2025,ruiz2025tunable}. Transport and magnetization measurements show that the Curie temperature is moderately suppressed as the Fe fraction $x$ decreases and the Pd content $y$ increases. Therefore, this work establishes the emerging vdW ferromagnets Fe$_{x}$Pd$_{y}$Te$_2$ as a complex but tunable system in which the chemical composition ($x$ and $y$) can be tuned to control both the magnetism and crystal structure.

\section{Experimental details}

\subsection{Single crystal growth}

Single crystals of Fe$_x$Pd$_{y}$Te$_2$ ($x=0.9-1$ and $y=2.5-2.3$) were produced by the high-temperature solution growth method \cite{canfield2019} as follows. Small pieces of Fe (Alfa Aesar, 99.99\%), Pd powder (DOE stockpile, 99.9+\%), and Te (Alfa Aesar, 99.999+\%) were weighed in molar ratios of Fe$_n$Pd$_{62}$Te$_{38}$ ($n$ = 23 and 34), and loaded into a 2 ml alumina frit-disc Canfield crucible set (CCS) \cite{canfield2016, LSP}. The initial compositions were chosen to allow the exploration of the low-melting regions near the Pd-Te eutectic (with approximate molar ratio 62:38 Pd:Te and eutectic temperature of $\sim 510\,^\circ$C) \cite{slade2022}. The packed CCS was flame-sealed into fused silica ampules that were backfilled with $\approx 1/6$ atm. Ar gas. Using a box furnace, the ampules were warmed to 800$\,^\circ$C over 10 h and held at that temperature for 6 h. Then, the furnace was gradually cooled to 550\,$^\circ$C over $\approx$ 80 h. Upon reaching the desired temperature, the excess liquid phase was decanted in a centrifuge with metal cups and rotors \cite{canfield2019}. After cooling to room temperature, the ampules and CSS were opened to reveal clusters of metallic plate-like crystals with typical dimensions of $\approx 1$ mm, as shown in the right inset of Fig. \ref{fig:pxrd}. For a direct comparison of crystal structures and physical properties, we also grew FePd$_2$Te$_2$ single crystals by following the synthetic protocol outlined in ref. \cite{shi2024}. The main processes are analogous to those described above, but the starting melt had composition FePd$_{2}$Te$_{2}$, and the samples were removed from the furnace at 500\,$^\circ$C, without any decanting step, where they were allowed to cool (quench) to room temperature. All of the Fe$_x$Pd$_y$Te$_2$ crystals we grew are air-stable over the length of at least 6--12 months, and may well be air-stable for much longer times.

\subsection{Elemental analysis}

The compositions of the crystals were determined by Energy Dispersive Spectroscopy (EDS) using a JEOL NeoScope JCM-7000 Benchtop scanning electron microscope (SEM), using an accelerating voltage of 15 kV. Quantitative analysis of EDS spectra were done with built-in software SMILE VIEW Map with factory standards. We analyzed four samples from each batch, and the composition of each crystal was measured at three or more spots, revealing good homogeneity in the cleavage plane. Additional details on EDS data are available in the supplementary information (SI), whereas the average EDS compositions are given in Table \ref{tab:eds} (first row).

\subsection{X-ray diffraction}

Room-temperature powder X-ray diffraction (PXRD) patterns were obtained using a Rigaku Miniflex-II instrument operating with Cu$-K\alpha$ radiation with $\lambda=1.5406\,$\AA $(K\alpha_1)$ and $1.5443\,$\AA $(K\alpha_2)$ at 30 kV and 15 mA. 

Single crystal X-ray diffraction (SCXRD) was performed using a Rigaku XtaLab Synergy-S diffractometer with Ag radiation ($\lambda = 0.56087$ \AA), in transmission mode, operating at 65 kV and 0.67 mA. The samples were held in a nylon loop with Dow Corning high vacuum grease, and the temperature was controlled using an Oxford Cryostream 1000. At least 20 minutes were waited at each temperature to ensure thermal stabilization before starting the measurement. The total number of collected runs and images was based on the strategy calculation from CrysAlisPro (Rigaku OD, 2023). Data integration and reduction were also performed using CrysAlisPro, and a numerical absorption correction was applied based on Gaussian integration over a face-indexed crystal. The high-temperature structures ($T = 400\,$K) were solved by intrinsic phasing using the SHELXT software package and refined with SHELXL, whereas the room-temperature structures were solved and refined with Jana2020, considering the main reflections only (i.e., disregarding the satellite peaks). 

\begin{table*}[htbp] 
\begin{ruledtabular}
\caption{Chemical compositions determined from EDS and SCXRD for Fe$_x$Pd$_{y}$Te$_2$ samples. The top row lists the nominal melt composition used for the solution growth. In columns 2 and 3, the melt compositions were Fe$_n$Pd$_{62}$Te$_{38}$ (see experimental details) labeled by the Fe fraction $n$. Data in the last column, labeled here as Fe$_{20}$Pd$_{40}$Te$_{40}$, refers to single crystals grown using a 1:2:2 ratio of Fe:Pd:Te following the method described by Shi et al. in ref. \cite{shi2024}.}
\label{tab:eds}
\begin{tabular}{llll}
     Melt Composition               & Fe$_{23}$Pd$_{62}$Te$_{38}$ ($n=23$)                                     & Fe$_{34}$Pd$_{62}$Te$_{38}$ ($n=34$)                                   & Fe$_{20}$Pd$_{40}$Te$_{40}$ (122)                         \\         \hline
EDS                  & Fe$_{0.94 (4)}$Pd$_{2.49 (2)}$Te$_2$      & Fe$_{1.05 (2)}$Pd$_{2.32 (2)}$Te$_2$     &  Fe$_{1.07 (4)}$Pd$_{1.82 (2)}$Te$_2$     \\
SCXRD ($T = 300\,$K) & Fe$_{0.93 (1)}$Pd$_{2.55 (1)}$Te$_2$      & Fe$_{1.09 (1)}$Pd$_{2.23 (1)}$Te$_2$     &  FePd$_{2.02 (2)}$Te$_2$                    \\      
SCXRD ($T = 400\,$K) & Fe$_{1.02 (2)}$Pd$_{2.32 (2)}$Te$_2$      & Fe$_{1.03 (1)}$Pd$_{2.33 (2)}$Te$_2$     &  FePd$_{2.04 (1)}$Te$_2$                    \\
\end{tabular}
\end{ruledtabular}
\end{table*}

\subsection{Physical property measurements}

Temperature-dependent resistance measurements were performed between 1.8 and 375\,K in a Quantum Design Physical Property Measurement System (PPMS), using Lake Shore AC resistance bridges (models 370 and 372), with a frequency of 16.2 Hz and a 3.16 mA excitation current. As the crystals naturally grew with a plate-like morphology, we measured the resistances along the cleavage plane. Contacts were made by spot welding a 25 $\mu$m thick annealed Pt wire onto the faces of the crystals in standard four-point geometry, yielding typical contact resistances of $\approx1\,\Omega$.

Magnetization measurements were performed in a Quantum Design Magnetic Property Measurement System (MPMS) SQUID magnetometer operating in the DC mode. The measurements were performed on cleaved samples with both the field applied in-plane and out-of-plane. For in-plane measurements, the samples were mounted between two straws. To measure out-of-plane, the samples were glued to a Kel-F disc, which was then placed inside a straw. The blank disc was first measured at the same temperatures and fields for background subtraction. Because the Kel-F disc is diamagnetic, some measurements show noise associated with zero crossing errors. Examples of these artifacts can be seen in Figure S8. High temperature magnetization was measured using the oven option of a Quantum Design MPMS3 instrument operating in VSM mode. In this case, no background subtraction was performed.

\section{Results and discussion}

\subsection{Chemical Composition}

Table \ref{tab:eds} shows the average chemical compositions determined from EDS measurements (first row). Additional details on the EDS analysis are given in the SI. In contrast with findings from Shi et al. \cite{shi2024}, our attempts to grow FePd$_2$Te$_2$ from direct combination reactions resulted in non-stoichiometric samples with approximate composition Fe$_{1.07}$Pd$_{1.82}$Te$_2$. We also obtained compounds with larger Pd fraction by changing the composition of the starting melt, nominally Fe$_n$Pd$_{62}$Te$_{38}$, with $n = 23$ producing samples with the approximate composition Fe$_{0.94}$Pd$_{2.49}$Te$_2$ and $n = 34$ resulting in Fe$_{1.05}$Pd$_{2.32}$Te$_2$. Whereas we did not exhaustively explore different starting compositions, these results indicate that FePd$_2$Te$_2$ has a substantial width of formation, and that the final composition of Fe$_x$Pd$_y$Te$_2$ single crystals can be controlled by changing the growth conditions. The refined chemical compositions, based on single crystal X-ray diffraction data (discussed in the following sections), reveal a similar width of formation; however, the refined Fe and Pd occupancies result in slightly different compositions from the EDS analysis. Owing to restrictions with our SCXRD refinements, which are detailed below, we consider the EDS compositions most reliable, and throughout the remainder of this manuscript, we will refer to the samples by their rounded EDS compositions Fe$_{1.1}$Pd$_{1.8}$Te$_2$, Fe$_{0.9}$Pd$_{2.5}$Te$_2$, and FePd$_{2.3}$Te$_2$, respectively, or for simplicity, their Pd fraction $y$, unless otherwise noted.

\subsection{Structural Characterization}

\begin{figure}[b]
    \centering
    \includegraphics[scale=1]{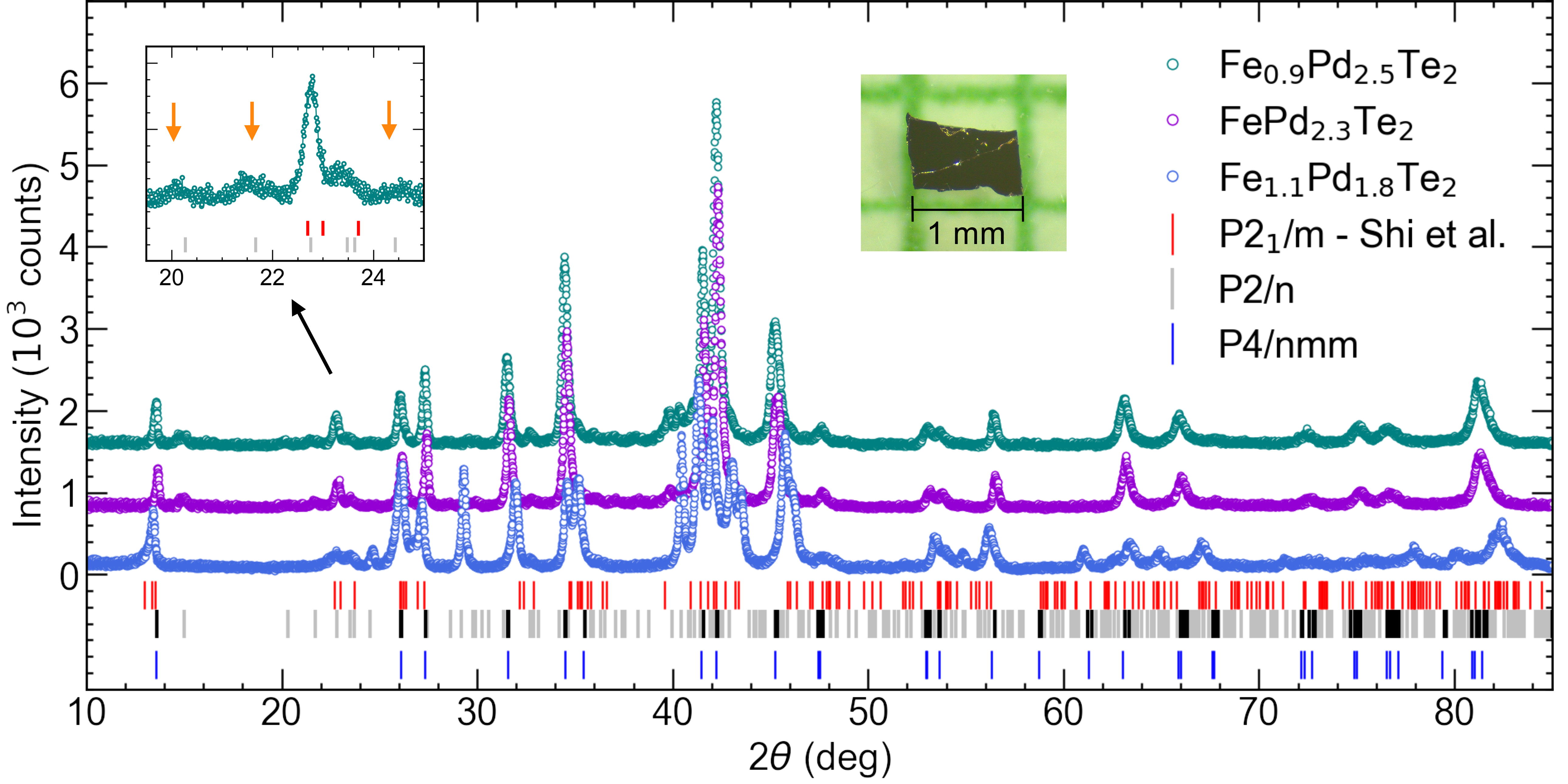}
    \caption{Room temperature powder X-ray diffraction patterns of Fe$_{x}$Pd$_{y}$Te$_2$. Black marks indicate the main reflections of the low-temperature phase (P$2/n$) of $y>2$ compounds, while gray marks stand for satellite peaks up to the second order. Peak positions of FePd$_2$Te$_2$ corresponding to the structure reported by Shi et al. \cite{shi2024} are shown in red. For comparison, main reflections of the high-temperature phase (P$4/nmm$) of $y>2$ compounds are also presented. The left inset shows a close-up of Fe$_{0.9}$Pd$_{2.5}$Te$_2$ data around $2\theta\approx22^\circ$, with orange arrows indicating the satellite peaks unaccounted for in the $P2_1/m$ structural model. The inset on the right shows a photo of a typical Fe$_{x}$Pd$_{y}$Te$_2$ single crystal.}
    \label{fig:pxrd}
\end{figure}

\begin{figure*}[t]
    \centering
    \includegraphics{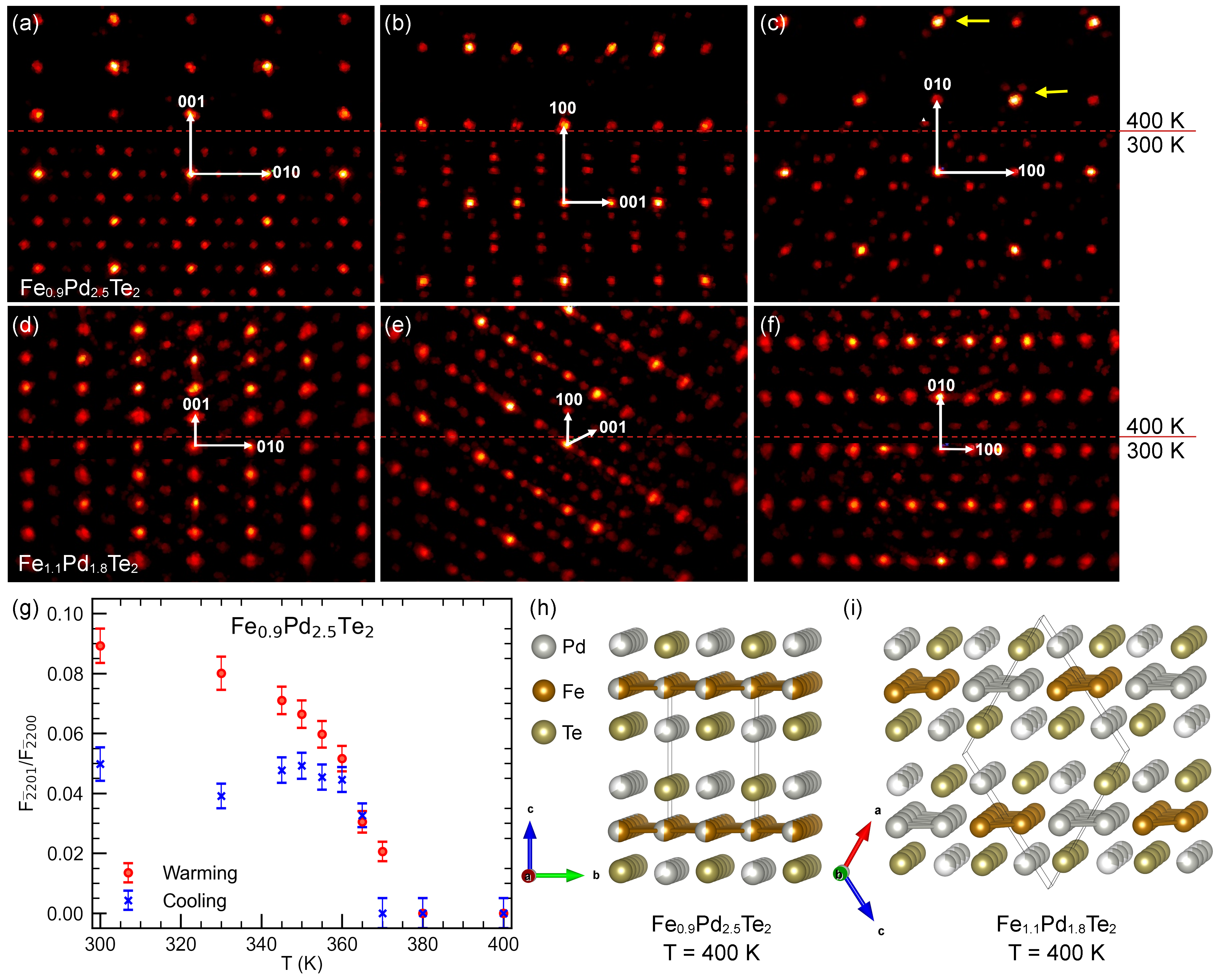}
    \caption{Single crystal diffraction of Fe$_{0.9}$Pd$_{2.5}$Te$_2$ along (a) 100, (b) 010, and (c) 001 directions. Additional peaks at 300 K signal a structural modulation with wave vector $q \approx \big(-0.42\,\,0.17\,\,\frac{1}{2}\big)$. Yellow arrows in (c) indicate peaks that split above the transition. Diffraction patterns for Fe$_{1.1}$Pd$_{1.8}$Te$_2$ along (d) 100, (e) 010, and (f) 001 directions, featuring extra intensities at $q = \big(\frac{1}{2}\,\,\frac{1}{2}\,\,\frac{1}{2}\big)$. (g) Integrated intensity of the first-order satellite peak associated with the main reflection $(\bar{2}20)$ upon warming and cooling for Fe$_{0.9}$Pd$_{2.5}$Te$_2$. The fourth index on the $y$-axis refers to the order of the satellite peak, with zero standing for a main reflection. (h) Undistorted tetragonal structure of Fe$_{0.9}$Pd$_{2.5}$Te$_2$ along (100) direction, emphasizing the layered structure along the \textit{c}-axis. (i) Average structure of Fe$_{1.1}$Pd$_{1.8}$Te$_2$ at 400\,K. Alternating Fe and Pd chains are seen in the cleavage plane. Unit cells are shown in black lines.}
    \label{fig:figure1}
\end{figure*}

Powder X-ray diffraction patterns obtained on Fe$_x$Pd$_y$Te$_2$ samples are shown in Fig. \ref{fig:pxrd}. Despite its non-stoichiometric composition, the pattern of Fe$_{1.1}$Pd$_{1.8}$Te$_2$ is consistent with the FePd$_2$Te$_2$ phase reported by Shi et al. \cite{shi2024}. Likewise, most of the peaks observed for the $y>2$ compounds are also consistent with Fe$_{1.1}$Pd$_{1.8}$Te$_2$ sample; however, as emphasized in the inset to Fig. \ref{fig:pxrd}, these patterns show weak additional reflections that cannot be indexed to the FePd$_2$Te$_2$ phase or binary impurities containing Fe, Pd, and/or Te. These results imply that the crystal symmetry of $y$ = 2.3 and $y$ = 2.5 compounds is different than that of the $y$ = 1.8 sample, suggesting that the crystal structure may be sensitive to chemical composition ($x$ and/or $y$ values). 

To explore this possibility further, we collected single crystal X-ray diffraction data on each sample. The bottom panels of Figs. \ref{fig:figure1}(a-c) and \ref{fig:figure1}(d-f) show (0$kl$), ($h$0$l$), and ($hk$0) planes for Fe$_{0.9}$Pd$_{2.5}$Te$_2$ and Fe$_{1.1}$Pd$_{1.8}$Te$_2$, respectively.  The most striking feature in the 300 K data is the clear presence of strong, but different, satellite peaks for $y$ = 1.8 and $y$ = 2.5 samples, implying that these compositions must have different crystal symmetries at room temperature. Next, we collected diffraction data at different temperatures from 300 to 400\,K, and found that for $y$ = 2.5, the satellite peak intensities steadily weaken on warming, completely vanishing above $\approx$ 370\,K. Diffraction patterns collected at 400\,K, shown in the top row of Figs. \ref{fig:figure1}(a-c), do not show any satellite peaks. As shown in Fig. \ref{fig:figure1}(g), upon cooling, the satellites reappear below $\approx$ 360\,K. Similar behavior was observed for $y=2.3$ (data available in Figure S2 of the SI). The appearance/disappearance of satellite peaks is reproducible on subsequent heating/cooling cycles and in separate samples, marking the presence of a structural phase transition near $\approx$ 370 K for Fe$_x$Pd$_y$Te$_2$ compounds with $y$ = 2.3 and 2.5. The 10\,K hysteresis associated with the appearance/disappearance of the satellites upon warming/cooling suggests that the transition is first-order in nature. This is supported by the observation of hysteresis on heating/cooling sweeps in magnetic and resistance measurements, as outlined in the discussion of the physical properties.

We also note that a close inspection of the $y$ = 2.5 diffraction patterns above the transition ($T = 400\,$ K) reveals additional intensities near the main reflections, for example, see the peaks indicated by yellow arrows in Fig. \ref{fig:figure1}(c). As outlined in the SI, these features are likely associated with misaligned domains that form when cycling the sample through the first-order structural transition. The appearance of these additional features is irreversible on heating/cooling up to 400 K; however, the extra intensities can be eliminated by annealing the samples at a higher temperature of 673 K (see Figure S6). The substantial difference in the intensities of the satellite peaks upon warming and cooling cycles, shown in Fig. \ref{fig:figure1}(g), is also likely associated with the formation of structural domains, which reduce the diffracted volume.

Diffraction data collected at 400\,K for Fe$_{1.1}$Pd$_{1.8}$Te$_2$ are shown in the top row of Figs. \ref{fig:figure1}(d-f) and are very similar to the respective room temperature data for this compound, suggesting that the structure has not changed at 400 K. As we will discuss below, magnetization measurements on this sample do show clear evidence for a transition at $\approx$ 420 K, but the maximum temperature of our diffractometer (400 K) restricts us from assessing its high-temperature structure. Importantly, the initial report on the discovery of FePd$_2$Te$_2$ did not observe satellite peaks \cite{shi2024}. We note that the satellite peaks observed here for Fe$_{1.1}$Pd$_{1.8}$Te$_2$ are very weak (for instance, $F_{\bar{2}201}/F_{\bar{2}200}\approx10^{-3}$ at 400\,K), and this may explain why they may have been missed in the earlier publication.

Figs. \ref{fig:figure1}(h) and \ref{fig:figure1}(i) illustrate the 400 K crystal structures determined from our SCXRD diffraction data for samples with respective compositions Fe$_{0.9}$Pd$_{2.5}$Te$_2$ and Fe$_{1.1}$Pd$_{1.8}$Te$_2$. Detailed information on the refinements is presented in Tables S1--S3 in the supplemental information, and Tables S4 and S5 list the refined atomic positions, site occupancies, and thermal displacement parameters corresponding to each composition at 300 K and 400 K, respectively.

From our diffraction data at $400\,$K, we find that in the high-temperature phase ($T>370\,$K), the \textit{y} = 2.3 and \textit{y} = 2.5  compounds adopt a tetragonal structure with space group P$4/nmm$ (n. 129), which is a disordered ternary derivative of the FeTe prototype. The crystal structure, shown  in Fig. \ref{fig:figure1}(h), consists of layered motifs, in which each layer is built from a central sheet of metal atoms (here disordered Fe and Pd) that form a square lattice in which the metal atoms are tetrahedrally coordinated to Te. The Fe and Pd atoms occupy the central (2a) position in each Te--M--Te (M = Fe/Pd) layer in an approximately 50:50 ratio. The Te--M--Te layers stack along the \textit{c}-axis, with a small vdW gap of $\approx$ 2.68 \AA. The remaining Pd is accommodated in the interstitial positions between Te atoms in the vdW gaps (Pd2 in Table S2). As noted in the experimental details, the Fe$_x$Pd$_{y}$Te$_2$ crystals form as plates, as illustrated by a representative sample in the right inset of Fig. \ref{fig:pxrd}. These plates can be cleaved with a razor blade, consistent with a 2D layered/vdW structure.

\begin{figure*}[t]
    \centering
    \includegraphics{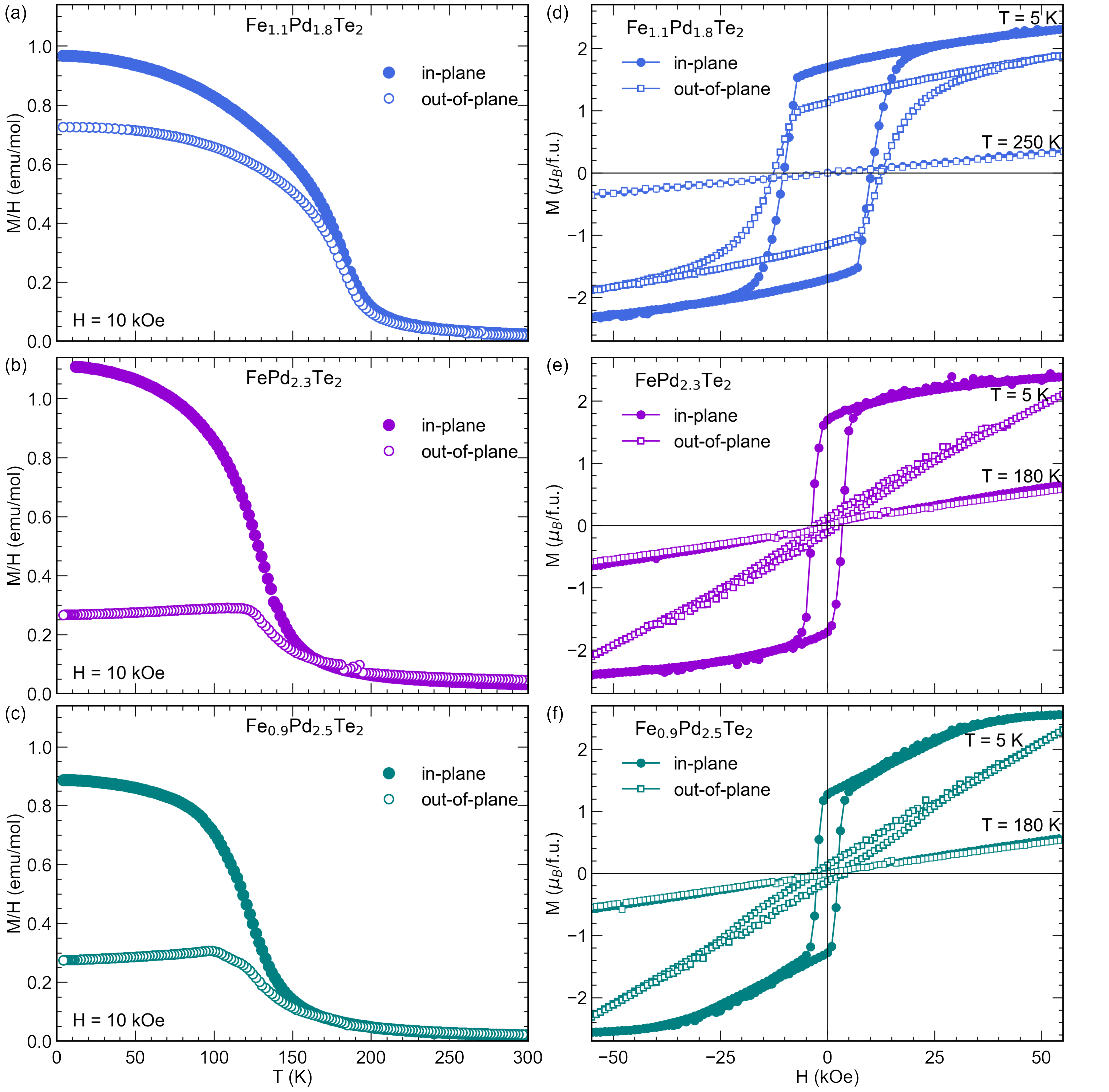}
    \caption{Anisotropic magnetization data for Fe$_x$Pd$_{y}$Te$_2$ single crystals ($x = 0.9-1.1$ and $y=2.5-1.8$). In-plane and out-of-plane measurements are represented by closed and open symbols, respectively. The in-plane orientation is $H\perp c$ for $y>2$ compounds and $H\perp [10\bar{1}]$ for Fe$_{1.1}$Pd$_{1.8}$Te$_2$, respectively. Temperature dependence of $M/H$ for (a) Fe$_{1.1}$Pd$_{1.8}$Te$_2$, (b) FePd$_{2.3}$Te$_2$, and (c) Fe$_{0.9}$Pd$_{2.5}$Te$_2$. These measurements were field-cooled with $H = 10\,$kOe. The noise around 180\,K in the out-of-plane measurements for $y>2$ compounds is from zero-crossing artifacts associated with the background subtraction (see experimental details). Field-dependent magnetization isotherms measured at $T = 5\,$K (circles) and $T = 180\,$K or $T=250\,$K (squares) for (d) $y$ = 1.8, (e) $y$ = 2.3, and (f) $y$ = 2.5 samples, respectively.} 
    \label{fig:figure3}
\end{figure*}

For $y$ = 2.3 and $y$ = 2.5 compositions, we observe both a change in crystal symmetry and the appearance of satellite peaks on cooling below the structural transition. If we neglect the satellite peaks and only integrate the main reflections, the average structure of the low-temperature phase can be described as a very subtle distortion of the tetragonal lattice, resulting in monoclinic P$2/n$ (n. 13) symmetry with $a = b  < c$, $\alpha$ = $\gamma= 90^\circ$, and $\beta\approx90.1^\circ$. The appearance of satellite peaks implies additional modulation of this average structure, and we find that, for both $y=2.3$ and $y$ = 2.5, a wave vector $q \approx \big(-0.42\,\,0.17\,\,\frac{1}{2}\big)$ provides a satisfactory description of all satellite peak positions at 300 K, and also accounts for the weak PXRD reflections shown in the inset of Fig. \ref{fig:pxrd}. Considering the high-temperature structure described above, the modulation in the low-temperature phase likely reflects the partial ordering of Fe/Pd atoms in the metal layer and/or the occupancy of the interstitial Pd atoms. Unfortunately, given the complex, highly disordered nature of Fe$_x$Pd$_{y}$Te$_2$, we have not yet found a model that fully describes the incommensurately modulated low-temperature structure. 

Nevertheless, useful information can still be inferred from the average structure that was arrived at by neglecting the satellite peaks, although we emphasize that features dependent on peak intensities, such as occupancies and corresponding chemical composition, should be taken with caution. In both low- and high-temperature phases of Fe$_{0.9}$Pd$_{2.5}$Te$_2$ and FePd$_{2.3}$Te$_2$, the disordered Fe/Pd atoms at the center of each Te--M--Te slab form zigzag chains. Above 370 K, the chains are equivalent in both \textit{a} and \textit{b} directions with the M--M--M (M = Fe or Pd) bonding angle equal to $90^\circ$ and the intrachain M--M distance about $2.83$ \AA, such that the metal atoms form a square lattice. As the temperature is decreased below the transition, symmetry is broken in the \textit{ab}-plane, resulting in nonequivalent intrachain M--M bonds, while the Te layers remain nearly unchanged. For example, in the average structure of Fe$_{0.9}$Pd$_{2.5}$Te$_2$ at 300\,K, the subtle distortion of the average structure results in intrachain M-M distances of $2.8339$\,\AA\, and $2.8411$\,\AA\, in [110] and [$\bar{1}10$] directions, respectively, corresponding to M-M-M bonding angles smaller than $90^\circ$. 

For Fe$_{1.1}$Pd$_{1.8}$Te$_2$, the main reflections (both at 300\,K and 400\,K) are well described by $P2_1/m$ symmetry, consistent with the results of Shi et al \cite{shi2024}. However, our data in Fig. \ref{fig:figure1}(d-f) also shows weak satellite peaks that were not observed in the initial publication and which remain observable up to at least 400 K. These satellite peaks can be indexed with a commensurate modulation wave vector of $q=\big(\frac{1}{2}\,\,\frac{1}{2}\,\,\frac{1}{2}\big)$. Without considering the satellite peaks, our structural solution and refined chemical composition (Table \ref{tab:eds}) agree well with the stoichiometric FePd$_2$Te$_2$.\cite{shi2024} The average structure can be visualized as a site-ordered derivative of the FeTe structure, see Fig. \ref{fig:figure1}(i). In contrast to the structure found in the compounds with the higher Pd content ($y>2$), the central layer of metal atoms is composed of ordered Fe/Pd sites, resulting in alternating Fe-Fe and Pd-Pd zigzag chains in the basal plane. The Te layer is composed of two nonequivalent $2e$ sites, leading to two nonequivalent interstitial positions partially occupied by Pd (Pd2 and Pd3 in Table S4). We emphasize again that the structural solutions at 300\,K and 400\,K for Fe$_{1.1}$Pd$_{1.8}$Te$_2$ were based only on the main reflections, and by neglecting the satellite peaks, the reliability of the refined site occupancy and thermal displacement parameters is questionable. Therefore, as discussed for the $y>2$ compounds, we attributed the difference between SCXRD and EDS compositions to this incomplete description of the diffraction data.

The most important finding from our diffraction data is the clear identification of a symmetry-breaking phase transition that occurs on cooling/heating around $\approx$ 370 K for the $y$ = 2.3--2.5 compounds and the indication (from superlattice peaks and magnetization data shown below) that a similar transition to a tetragonal structure may occur for Fe$_{1.1}$Pd$_{1.8}$Te$_2$ above 400\,K. Furthermore, STM data from Shi et al. and polarized optical microscopy images in more recent publications show that FePd$_2$Te$_2$ contains rich, sub-micron scale, structural twins with a 90$^\circ$ relative orientation between domains \cite{shi2024, chen2024, mi2025,ruiz2025tunable}. The origin of the crystal twinning remains a key open question left by previous works, motivated by the possibility that the macroscopic spin texture associated with these domains may support interesting magnetotransport behavior. Symmetry breaking from a high-temperature tetragonal structure, as occurs when cooling through the structural transition identified here, provides a very strong explanation for the origin of these structural domains.

\subsection{Magnetic and Transport Properties}

Field-cooled temperature-dependent magnetization $M/H$ collected at $H = 10\,$kOe for Fe$_x$Pd$_{y}$Te$_2$ single crystals are presented in Figs. \ref{fig:figure3}(a-c). The in-plane orientation is the \textit{ab}-plane ($H\perp c$) for $y>2$. For the Fe$_{1.1}$Pd$_{1.8}$Te$_2$ sample, with space group P$2_1/m$ , the in-plane direction is $H\perp [10\bar{1}]$. Ferromagnetic ordering is observed when applying the field both in-plane and out-of-plane. Whereas the in-plane $M/H$ is larger than the out-of-plane values for Fe$_{1.1}$Pd$_{1.8}$Te$_2$, we observe a substantially more anisotropic response for $y=2.3$ and 2.5 samples, with the $M/H$ results indicating that the in-plane orientation is the easy direction.

Above the respective Curie temperatures, the \textit{M}(\textit{H}) isotherms displayed in Figures \ref{fig:figure3}(d-f) are essentially linear over the full field range, consistent with paramagnetic behavior. We note that a close-up view of the data below 5 kOe (shown in Figure S8 in the SI) reveals non-linear behavior at low fields, likely suggesting the samples contain a very small quantity of a $T_C >$ 250 K  ferromagnetic impurity. The \textit{T} = 5 K \textit{M}(\textit{H}) isotherms are consistent with the temperature-dependent data. For \textit{y} = 2.3--2.5, we observed moderate hysteresis with coercive fields of $\approx$ 2 kOe when the field is applied in the basal plane. The out-of-plane \textit{M}(\textit{H}) isotherms are nearly linear, where the weakly hysteretic behavior likely reflects slight sample misorientation. These results confirm that \textit{y} = 2.3--2.5 Fe$_x$Pd$_{y}$Te$_2$ samples are moderately hard ferromagnets with moments aligned within the basal plane. On the other hand, isotherms for Fe$_{1.1}$Pd$_{1.8}$Pd$_2$ have substantial hysteresis, with coercive fields near 10 kOe, for both in-plane and out-of-plane orientations, as shown in Fig. \ref{fig:figure3}(d). Our magnetic data for the Fe$_{1.1}$Pd$_{1.8}$Te$_2$ composition is very similar to those previously reported for FePd$_2$Te$_2$ \cite{shi2024}, with the strong, comparably isotropic, coercivity implying that the ferromagnetic moments either have both in-plane and out-of-plane components for $y=1.8-2$, or the pinning of FM domains is substantially stronger in this sample. Neutron diffraction will clearly be needed to clarify the magnetic structure of these compounds.

\begin{figure}[t]
    \centering
    \includegraphics{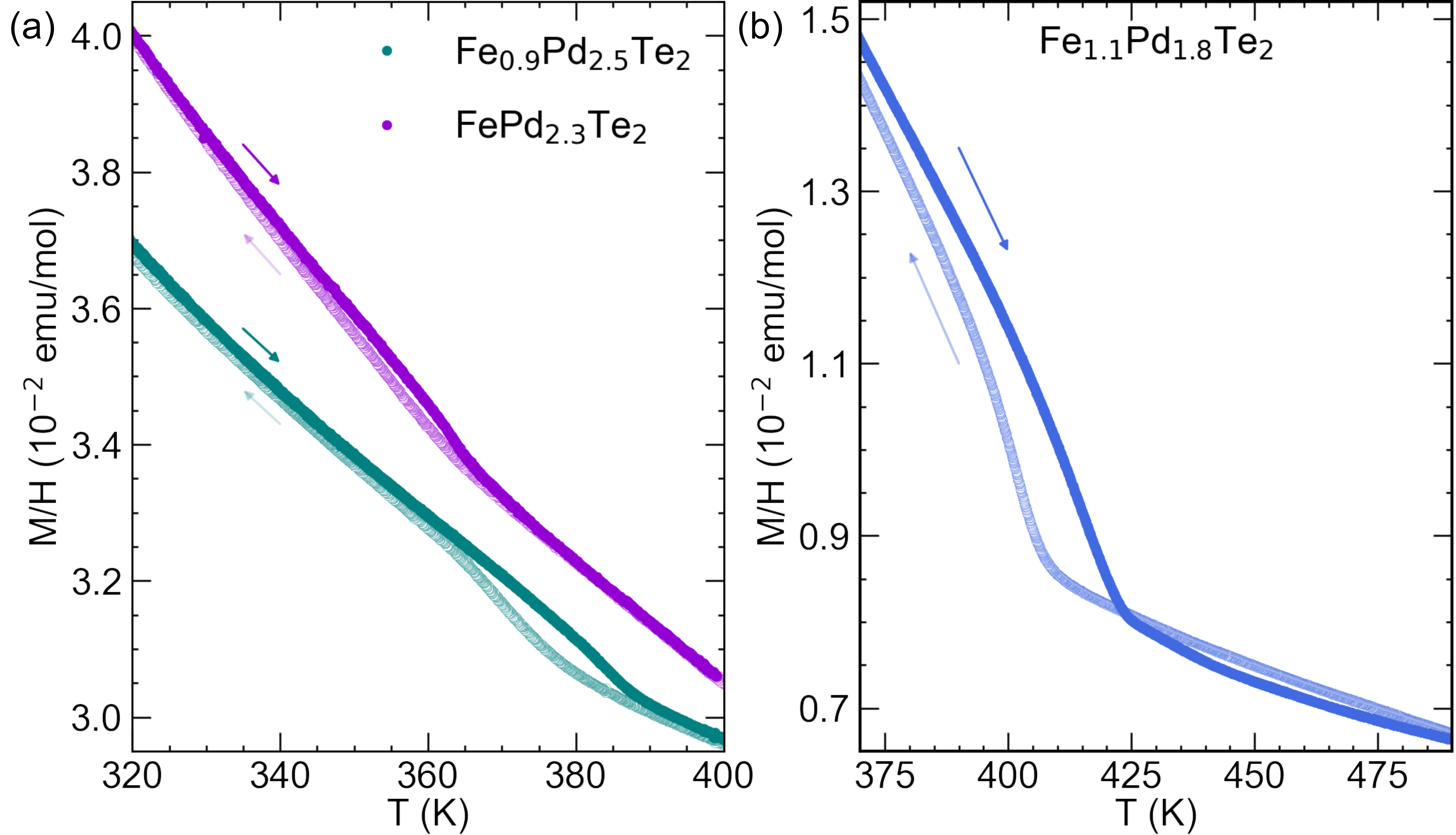}
    \caption{High-temperature $M/H$ for Fe$_x$Pd$_{y}$Te$_2$ single crystals with (a) $y > 2$ and (b) $y = 1.8$. All measurements were field-cooled with $H = 10\,$kOe applied in-plane. All compounds exhibit hysteresis on increasing (solid circles) and decreasing (open circles) temperature sweeps around the phase transition.}
    \label{fig:highT}
\end{figure}  

\begin{table*}[t] 
\begin{ruledtabular}
\caption{In-plane and out-of-plane magnetic properties of Fe$_x$Pd$_{y}$Te$_2$ single crystals. The coercive field $H_c$ and saturated moment $\mu_{\text{sat}}$ were estimated at 5\,K. The magnetization at 55\,kOe was taken as an estimate of $\mu_{\text{sat}}$. Weiss temperature $\Theta$ and effective moment $\mu_{eff}$  were obtained from CW fits of the high-temperature paramagnetic susceptibility. Curie temperatures $T_C$ inferred from resistance data, shown in Fig. \ref{fig:figure2}(a), are also given. Uncertainties estimated from linear or CW fits are given in parentheses. The considerably larger uncertainties of $\Theta$ values in out-of-plane measurements are due to the noise arising during background subtraction, see Figure S8.}
\label{tab:magproperties}
\begin{tabular}{cccccc|cc}
\multicolumn{6}{c|}{In-plane}                                                                                                       & \multicolumn{2}{c}{Out-of-plane}         \\ \hline
Compound                   & $T_{C}$ (K) & $\Theta$ (K) & $\mu_{eff}$ ($\mu_B$/f.u) & $\mu_{\text{sat}}$ ($\mu_B$/f.u) & $H_c$ (kOe) & $\Theta$ (K) & $\mu_{eff}$ ($\mu_B$/f.u) \\ \hline
Fe$_{1.1}$Pd$_{1.8}$Te$_2$ & 180         & 172.7 (2)    & 4.93 (1)                  & 2.31(1)                & 10          & 155 (67)     & 5.8 (4)                   \\
FePd$_{2.3}$Te$_2$         & 112         & 125.9 (2)    & 5.65 (1)                  & 2.39(1)                & 3.8         & 121 (8)      & 5.48 (6)                  \\
Fe$_{0.9}$Pd$_{2.5}$Te$_2$ & 98          & 121.2 (6)    & 5.74 (1)                  & 2.56(1)                & 2.5         & 121 (4)      & 5.39 (2)                    
\end{tabular}
\end{ruledtabular}
\end{table*}

In addition to the ferromagnetic order, the in-plane magnetization of all Fe$_{x}$Pd$_{y}$Te$_2$ compounds exhibit an anomaly above room temperature, as shown in Fig. \ref{fig:highT}. Upon warming, the magnetization clearly changes its slope at $\approx$ 365 K and $\approx$ 385 K for $y=2.3$ and 2.5, respectively, in reasonable agreement with the temperatures at which the structural change occurred in our X-ray diffraction data. The nearly 10 K hysteresis observed in Fig. \ref{fig:highT}(a) is also consistent with the diffraction data, see Fig. \ref{fig:figure1}(g), and confirms that the transition is first order. A stronger, but similar, feature is observed at $\approx$ 420 K for $y=1.8$, as shown Fig. \ref{fig:highT}(b). Given the presence of satellite peaks in the 300-400 K diffraction data for this compound, the high-temperature anomaly observed in magnetization strongly suggests that Fe$_{1.1}$Pd$_{1.8}$Te$_2$ also undergoes a structural phase transition, and based on the minima in the derivative of the $M/H$ data collected upon warming, we estimated the transition temperature as 417\,K. Considering the similar behavior observed across the Fe$_x$Pd$_{y}$Te$_2$ series, we speculate that the Fe$_{1.1}$Pd$_{1.8}$Te$_2$ composition also adopts a tetragonal FeTe-derived structure above 417 K; however, \textit{T} $>$ 420 K diffraction measurements are needed to confirm its high-temperature structure. We emphasize again that the identification of this first-order structural transition in Fe$_{1.1}$Pd$_{1.8}$Te$_2$ may explain the origin of the 90$^\circ$ structural twins observed in previous works \cite{shi2024,chen2024, mi2025}.

Table \ref{tab:magproperties} summarizes the magnetic properties of the Fe$_x$Pd$_{y}$Te$_2$ single crystals. Details on the Curie-Weiss fits to the high-temperature susceptibility, which were used to estimate the effective magnetic moments ($\mu_{\text{eff}}$) and Weiss temperatures ($\Theta$), are outlined in the supporting information. In particular, owing to the low field (\textit{H} $<$ 5 kOe) non-linearity in the \textit{M}(\textit{H}) isotherms, we estimated the susceptibility as $\chi\approx\Delta M/\Delta H$ with $\Delta M = M(\text{20 kOe}) - M(\text{10 kOe})$ and $\Delta H =\text{20 kOe} - \text{10 kOe}$, where the associated data is shown in Figures S7 and S8 of the SI. As the 5 K \textit{M}(\textit{H}) isotherms all are nearly saturated at our highest fields, we estimated the saturated moments, $\mu_{\text{sat}}$, as \textit{M}(55 kOe) at 5 K. These compounds are all relatively hard, planar ferromagnets, for which the Curie temperature, in-plane Weiss temperature, and coercive fields all decrease by increasing the Pd fraction $y$, which is accompanied by a decrease in the Fe fraction $x$. Our data on Fe$_{1.1}$Pd$_{1.8}$Te$_2$ agrees well with the values reported by Shi et al. for FePd$_2$Te$_2$ \cite{shi2024}. We also find that the $y>2$ compounds exhibit stronger in-plane anisotropy, likely reflecting the greater fraction of relatively high-\textit{Z} Pd atoms. Furthermore, we observe a small increase in both the effective ($\mu_{\text{eff}}$) and saturated ($\mu_{\text{sat}}$) moments as the samples become more Pd-rich, where the effective moments for $y$ = 2.3 and $y$ = 2.5 compositions are slightly larger than the theoretical value of 4.9\,$\mu_B$ expected for Fe$^{2+}$ ($S=2$) moments. These trends may suggest an unquenched orbital contribution or partial Pd polarization, the latter of which is consistent with our observation of larger moments in the Pd-rich samples. 

\begin{figure}[b]
    \centering
    \includegraphics{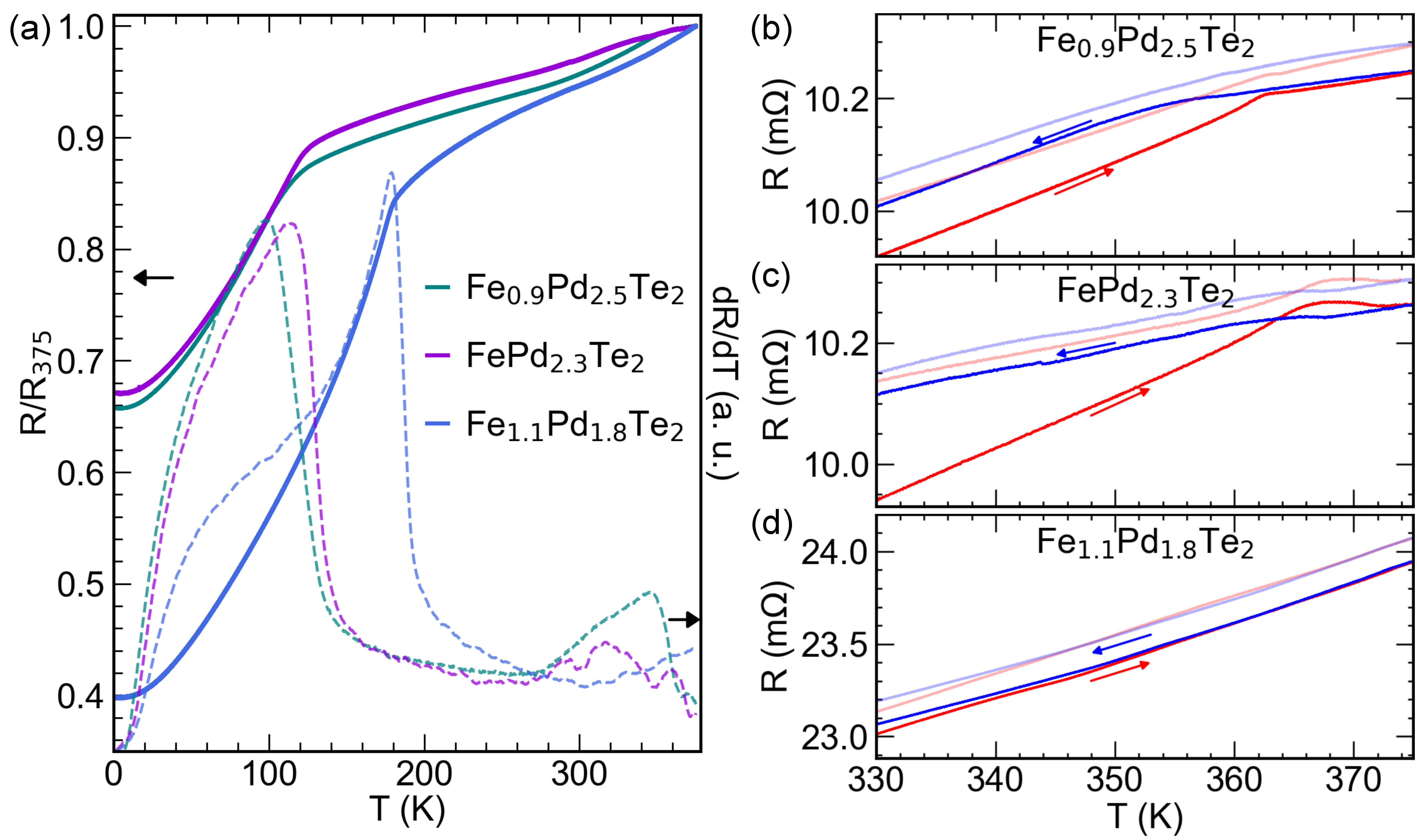}
    \caption{(a) Temperature-dependent resistance upon cooling normalized to its value at 375\,K (left axis), $R/R_{375}$, for Fe$_x$Pd$_{y}$Te$_2$ samples with $x = 0.9-1.1$ and corresponding $y=2.5-2.3,\, 1.8$. Resistance derivatives (right axis), $dR/dT$, are shown in dashed lines. Resistance above $T > 330\,$ K for (b) Fe$_{0.9}$Pd$_{2.5}$Te$_2$, (c) FePd$_{2.3}$Te$_2$ and (d) Fe$_{1.1}$Pd$_{1.8}$Te$_2$ measured upon warming and cooling (red and blue lines, respectively), emphasizing the hysteretic behavior of the phase transition around 360\,K for $y>2$ compounds. Shaded lines are measurements collected in a subsequent warming/cooling cycle.}
    \label{fig:figure2}
\end{figure}

The temperature-dependent resistance for Fe$_{x}$Pd$_{y}$Te$_2$ single crystals is shown in Fig. \ref{fig:figure2}. The residual resistance ratios, RRR $= R_{375}/R_{1.8}$, are low, around 1.5 for both $y=2.3$ and 2.5, consistent with the structural disorder inferred from our X-ray diffraction data. The RRR of Fe$_{1.1}$Pd$_{1.8}$Te$_2$ is a bit larger, about 2.5, as seen in the blue curve in Fig. \ref{fig:figure2}(a), in agreement with its less disordered structure, with distinct Fe and Pd sites in the metal layer. The resistance in all samples exhibits metallic temperature dependence that decreases on cooling, with a clear kink followed by a rapid drop of resistance around 120\,K for $y>2$ and at 180\,K for $y=1.8$, respectively. The observed kinks are a typical feature of magnetic ordering in metals, marking the loss of spin-disorder scattering as the samples enter a magnetically ordered state. Based on the maxima in the derivatives of the $R/R_{375}$ data \cite{fisher1968}, shown in dashed lines in Fig. \ref{fig:figure2}(a), we estimate the magnetic transition temperatures to be $T_C = 98\,$K, $112\,$K, and $180\,$K for $y$ = 2.5 $y$ = 2.3 and $y$ = 1.8, respectively. Therefore, a decrease in the Pd fraction $y$, and corresponding increase in the Fe fraction $x$ from 0.9 to 1.1, results in higher Curie temperatures. The value observed for Fe$_{1.1}$Pd$_{1.8}$Te$_2$ is in excellent agreement with the reported $T_C$ of FePd$_{2}$Te$_2$ \cite{shi2024}.

For $y$ = 2.3 and $y$ = 2.5 compositions, a second kink and slope change in the resistance is observed above $T>360\,$K, consistent with the structural change identified in both diffraction and magnetization data. Figures \ref{fig:figure2}(b, c) show a close-up view of the resistance between 330--375\,K. The resistance reaches a shallow minimum at the transition temperature, which is most apparent in the warming data for FePd$_{2.3}$Te$_2$, as seen in Fig. \ref{fig:figure2}(c). Like in the magnetic and diffraction measurements, the resistance is also hysteretic around the transition temperature, consistent with the first-order nature. Additionally, the shaded lines in Figures \ref{fig:figure2}(b-c) show that the resistance gradually increases and that the signature of the transition becomes weaker on subsequent warming/cooling cycles. We attribute this irreversibility to increased scattering associated with mismatched structural domains, which form when cycling the sample through the first-order transition (see detailed discussion in the SI), consistent with the irreversible changes observed in the SCXRD data when warming the samples above the transition temperature. 

We note that while roughly consistent, the resistive signatures of the structure change occur at somewhat different temperatures than those inferred from the magnetic data, especially for the $y$ = 2.5 sample. Furthermore, magnetic measurements on different crystals of the same composition (batch) show $\approx$ 10 K variation in transition temperature. It is possible that that these differences reflect the thermal hysteresis observed near the transition. An alternative explanation is that the transition is sensitive to strains associated with differential thermal expansion between the sample, an exfoliable vdW material, and the sample holder. At the present, we do not fully understand the discrepancy; however, we emphasize that this does not change any of the key conclusions of this manuscript: that a transition from an incommensurately modulated to disordered FeTe-like tetragonal structure occurs near \textit{T} $\approx$ 370 K. Finally, Figure \ref{fig:figure2}(d) shows that the resistance of the composition Fe$_{1.1}$Pd$_{1.8}$Te$_2$ presents no feature between 300-375 K, in agreement with our magnetic measurements that show the transition in this sample occurs near 420 K.

\section{Summary and conclusions}


Single crystals of Fe$_x$Pd$_{y}$Te$_2$ were synthesized with $x$ ranging from 0.9 to 1.1 and $y$ from 1.8 to 2.5, respectively. The Fe$_x$Pd$_{y}$Te$_2$ samples are air-stable milliliter-scale plate-like single crystals, which can be readily cleaved with a razor blade. All compounds in the Fe$_x$Pd$_{y}$Te$_2$ series are metallic, easy-plane ferromagnets with magnetic properties, such as the Curie temperatures and coercive field, varying with Fe and Pd concentrations, $x$ and $y$, which can be moderately controlled by adjusting the growth conditions. In particular, the $y>2$ compounds show much stronger in-plane anisotropy than the $y$ = 1.8 composition. For $y$ = 2.3 and $y$ = 2.5, we identified a first-order structural phase transition at \textit{T} $\approx$  370 K, which was observed in diffraction, magnetic, and transport measurements. In the high-temperature phase, these systems adopt a derivative of the layered tetragonal P$4/nmm$ FeTe-structure featuring heavy site occupancy/interstitial disorder. In the low-temperature phase, the tetragonal symmetry is broken, and the room temperature structures for $y$ = 2.3 and $y$ = 2.5 compositions are incommensurately modulated with wave vector $q$ $\approx$ $\big(-0.42\,\,0.17\,\,\frac{1}{2}\big)$, where the modulation is likely associated with the complex ordering of the interstitial Pd and/or the Fe/Pd atoms in the central layers. 

Our crystallographic, transport, and magnetic characterizations of the Fe$_{1.1}$Pd$_{1.8}$Te$_2$ composition are very similar to those reported for FePd$_2$Te$_2$ by Shi et al. Notably, however, we observed $q=\big(\frac{1}{2}\,\,\frac{1}{2}\,\,\frac{1}{2}\big)$ satellite peaks in our diffraction data, implying additional modulation of the monoclinic structure previously reported. Furthermore, magnetization measurements also show an anomaly at $\approx$ 420 K, similar to the features observed in $y>2$ samples. Therefore, we speculate that the Fe$_{1.1}$Pd$_{1.8}$Te$_2$ compound also adopts a tetragonal structure above 420 K. Previous studies of FePd$_2$Te$_2$ noted the presence of perpendicularly oriented, sub-micron structural domains. Here, the observation of a symmetry-breaking structural transition from a parent tetragonal structure establishes a likely explanation for the origin of these domains.  

Further experimental investigations are necessary to determine the high-temperature structure of Fe$_{1.1}$Pd$_{1.8}$Te$_2$ and to fully describe the structural modulation observed in the low-temperature phase of all members of the Fe$_x$Pd$_{y}$Te$_2$ series. It is unclear what mechanisms are involved in the change from commensurate to incommensurate modulation and how this relates to the Fe and Pd fractions. Also, attempts to grow single crystals with fully occupied interstitial sites, resulting in FePd$_3$Te$_2$ chemical composition, have so far failed and may indicate a limit of stability of the distorted-tetragonal structure. Lastly, the individual role of $x$ and $y$ fractions in the magnetism of these materials are yet to be explored, with our results implying that these may be useful tunable parameters for the controlling and engineering of the physical properties of Fe$_x$Pd$_{y}$Te$_2$ single crystals.

\begin{acknowledgments}
Work at Ames National Laboratory was supported by the U.S. Department of Energy (DOE), Basic Energy Sciences, Division of Materials Sciences \& Engineering, under Contract No. DE-AC02-07CH11358. RFSP's one-year visit to Ames Laboratory and Iowa State University was supported by the São Paulo Research Foundation (FAPESP), Brasil, Process Number 2024/08497-6. SLM acknowledges support from FAPESP under Process Number 2023/10775-1.
\end{acknowledgments}

\bibliography{bib}

\end{document}


\title{Supplemental Information \\
\vskip 0.5cm 
Structural phase transitions in the van der Waals ferromagnets Fe$_x$Pd$_{y}$Te$_2$}

\author{R. F. S. Penacchio}
\affiliation{Ames National Laboratory, US DOE, Iowa State University, Ames, IA, USA}
\affiliation{Department of Physics, Iowa State University, Ames, Iowa 50011, USA}
\affiliation{Institute of Physics, University of S{\~{a}}o Paulo, S{\~{a}}o Paulo, SP, Brazil}

\author{S. Mohamed}
\affiliation{Ames National Laboratory, US DOE, Iowa State University, Ames, IA, USA}
\affiliation{Department of Chemistry, Iowa State University, Ames, Iowa 50011, USA}

\author{S{\'{e}}rgio L. Morelh{\~{a}}o}
\affiliation{Institute of Physics, University of S{\~{a}}o Paulo, S{\~{a}}o Paulo, SP, Brazil}

\author{S. L. Bud’ko}
\affiliation{Ames National Laboratory, US DOE, Iowa State University, Ames, IA, USA}
\affiliation{Department of Physics, Iowa State University, Ames, Iowa 50011, USA}

\author{P. C. Canfield}
\affiliation{Ames National Laboratory, US DOE, Iowa State University, Ames, IA, USA}
\affiliation{Department of Physics, Iowa State University, Ames, Iowa 50011, USA}

\author{T. J. Slade}
\affiliation{Ames National Laboratory, US DOE, Iowa State University, Ames, IA, USA}

\renewcommand{\thefigure}{S\arabic{figure}}
\setcounter{figure}{0}

\renewcommand{\thetable}{S\arabic{table}}
\setcounter{table}{0}

\maketitle
 
\clearpage
\onecolumngrid

\newpage

\section{EDS analysis}

\begin{figure*}[htbp]
    \centering
    \includegraphics[scale=1.3]{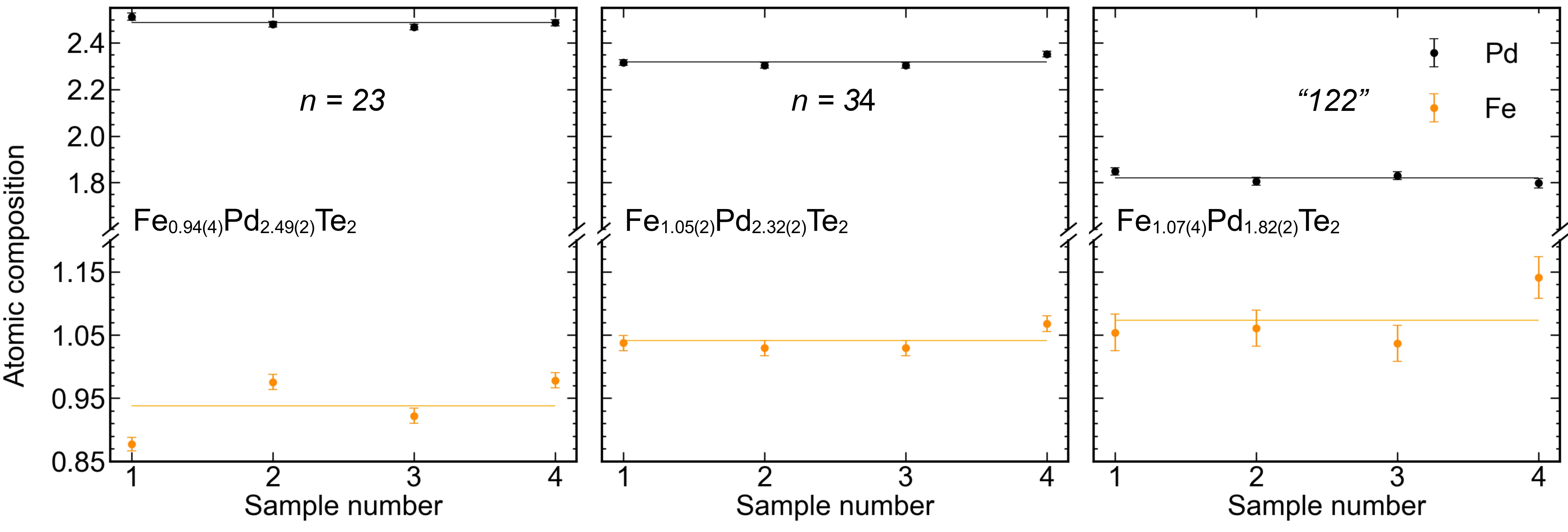}
    \caption{EDS data for Fe$_x$Pd$_y$Te$_2$ samples grown from melts with nominal compositions Fe$_n$Pd$_{62}$Te$_{38}$ (left and center) and Fe$_{20}$Pd$_{40}$Te$_{40}$ (right) in the starting melt. The label (122) here refers to the attempt of growing a stoichiometric compound FePd$_2$Te$_2$ following the protocol reported by Shi et al. \cite{shi2024} For each sample, spectra were collected at three different spots, yielding an average chemical composition with corresponding error bars. Batch average compositions are the average values obtained for the four samples (horizontal lines), with the corresponding standard deviation as uncertainties.} 
    \label{fig:figure1_SI}
\end{figure*}

For each batch with nominal starting composition Fe$_n$Pd$_{62}$Te$_{38}$ ($n= 23$ and 34) or Fe$_{20}$Pd$_{40}$Te$_{40}$ (i.e. FePd$_2$Te$_2$), the chemical composition of four samples were determined through energy dispersive X-ray spectroscopy (EDS). In our analysis, we forced the Te fraction to be 2, and the Fe and Pd fractions are given in comparison to Te. To evaluate sample homogeneity, each sample was measured at three different spots, yielding an average value with a corresponding error bar (standard deviation). Figure \ref{fig:figure1_SI} presents the results. Batch average compositions with corresponding uncertainties (between parentheses) are also shown. As pointed out in the main text, $n=23$ and 34 lead to single crystals of approximate compositions Fe$_{0.9}$Pd$_{2.5}$Te$_2$ and FePd$_{2.3}$Te$_2$, respectively. Our attempt to grow the stoichiometric FePd$_2$Te$_2$ compound, following the method described by Shi et al. \cite{shi2024}, resulted in  Fe$_{1.1}$Pd$_{1.8}$Te$_2$. Lastly, as seen from the small error bars, roughly $10^{-2}$, good sample homogeneity was observed in the cleavage plane for all samples.

\clearpage
\section{Additional diffraction details}

\begin{figure*}[b]
    \centering
    \includegraphics{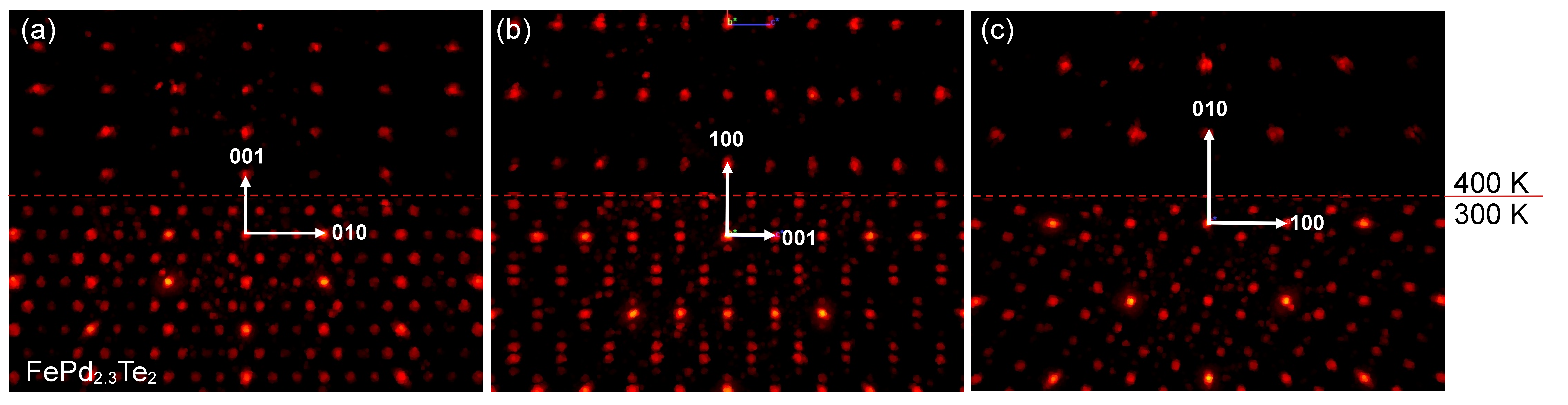}
    \caption{Single crystal diffraction of FePd$_{2.3}$Te$_2$ along (a) 100, (b) 010, and (c) 001 directions. Additional peaks at 300 K signal a structural modulation with wave vector $q \approx \big(-0.42\,\,0.17\,\,\frac{1}{2}\big)$.}
    \label{fig:zx832}
\end{figure*}

Single crystal diffraction collected at 300\,K and 400\,K for FePd$_{2.3}$Te$_2$ are shown in Fig. \ref{fig:zx832}. Like the data discussed in the main text for $y=2.5$, the patterns for FePd$_{2.3}$Te$_2$ show strong satellite peaks at 300 K that weaken on warming and vanish above $\approx$ 370 K, with no satellites being observed at 400 K. The room temperature modulation has wave vector $q\approx\big(-0.42\,\,0.17\,\,\frac{1}{2}\big)$, identical to the \textit{q}-vector observed for Fe$_{0.9}$Pd$_{2.5}$Te$_2$. 

Tables \ref{tab:refinementZX679}--\ref{tab:refinementAB065} provide information regarding the structural refinement of single crystal data for Fe$_x$Pd$_y$Te$_2$ samples, and Tables \ref{tab:structures} and \ref{tab:structures300K} list the atomic coordinates, site occupancies, and thermal displacement parameters in the 400 K and 300 K structures of each composition. As discussed in the main text, the structural complexity of these materials, with both positional and interstitial disorders, has thus far challenged our ability to satisfactorily solve the modulated structures below the transition. Owing to these difficulties, we emphasize here that our refinements considered only the average structure, i.e. the satellite peaks were neglected, so that structural parameters related to peak intensities (site occupancy and thermal displacement parameters) should be treated with caution.

\begin{figure*}
    \centering
    \includegraphics[scale=1.07]{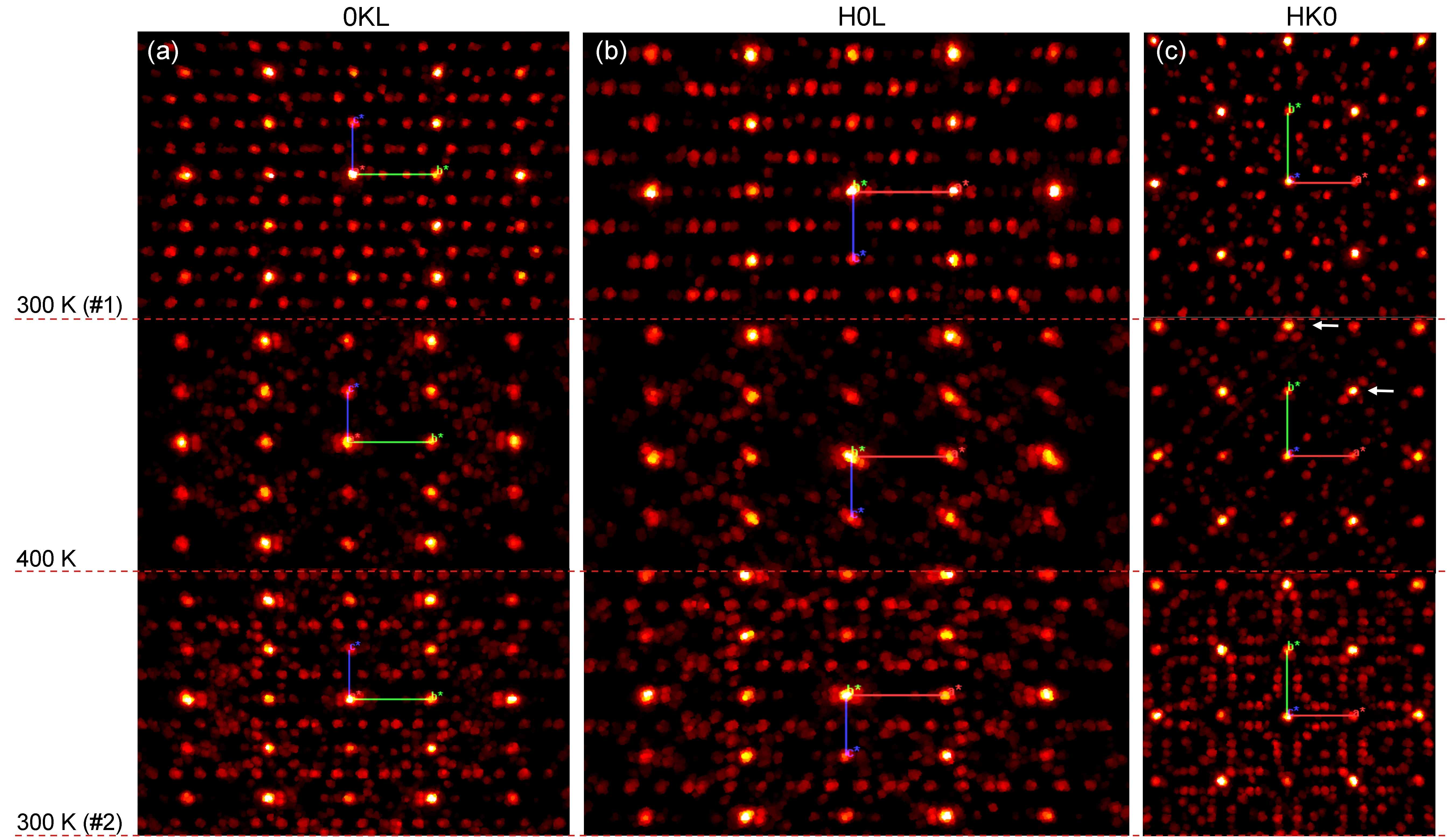}
    \caption{Single crystal diffraction of Fe$_{0.9}$Pd$_{2.5}$Te$_2$ along (a) 100, (b) 010, and (c) 001 directions during a warming/cooling cycle. First row shows the data collected at room temperature in an as-grown crystal -- 300\,K (\#1). Middle and bottom panels stand for subsequent data collections at 400\,K and 300\,K (\#2), respectively. White arrows in the high-temperature data (400\,K) at the HK0 direction indicate the (020) and (110) reflections, where the peak splitting is easily seen.} 
    \label{fig:SCXRD_SI}
\end{figure*}

As highlighted in Figure 2(a-c) of main text, single crystal diffraction patterns collected above the transition, ($T>370\,$K), for \textit{y} = 2.3 and 2.5 samples, exhibit extra features in addition to the reflections of the tetragonal phase. Figure \ref{fig:SCXRD_SI} shows the same regions of the reciprocal space upon warming/cooling but with a lower intensity threshold. Data collected at 300\,K (\#1) using an as-grown crystal, top panel of Fig. \ref{fig:SCXRD_SI}, exhibit only the main reflections and strong satellite peaks, which are associated with the complex incommensurate modulation discussed in the main text and above. As the temperature is increased above the transition, all satellite peaks vanish; however, we observe new features. Most prominently, the \{200\} and \{110\} reflections split into three peaks, as indicated by white arrows in Fig. \ref{fig:SCXRD_SI}(c) (middle panel). Also, a non-negligible intensity is distributed along the [110] direction. Interestingly, these features appear to be irreversible once the sample is heated through the transition, and do not vanish upon cooling, so that subsequent diffraction data collected at 300\,K, shown in the bottom line of Fig. \ref{fig:SCXRD_SI}, is an exceedingly complicated superposition of the satellite peaks described by $q\approx \big(-0.42\,\,0.17\,\,\frac{1}{2}\big)$ with these new peaks. This apparent peak splitting that occurs after warming above the transition temperature is reproducible and was observed in multiple measurements on different samples.

\begin{figure}
    \centering
    \includegraphics[scale=1.07]{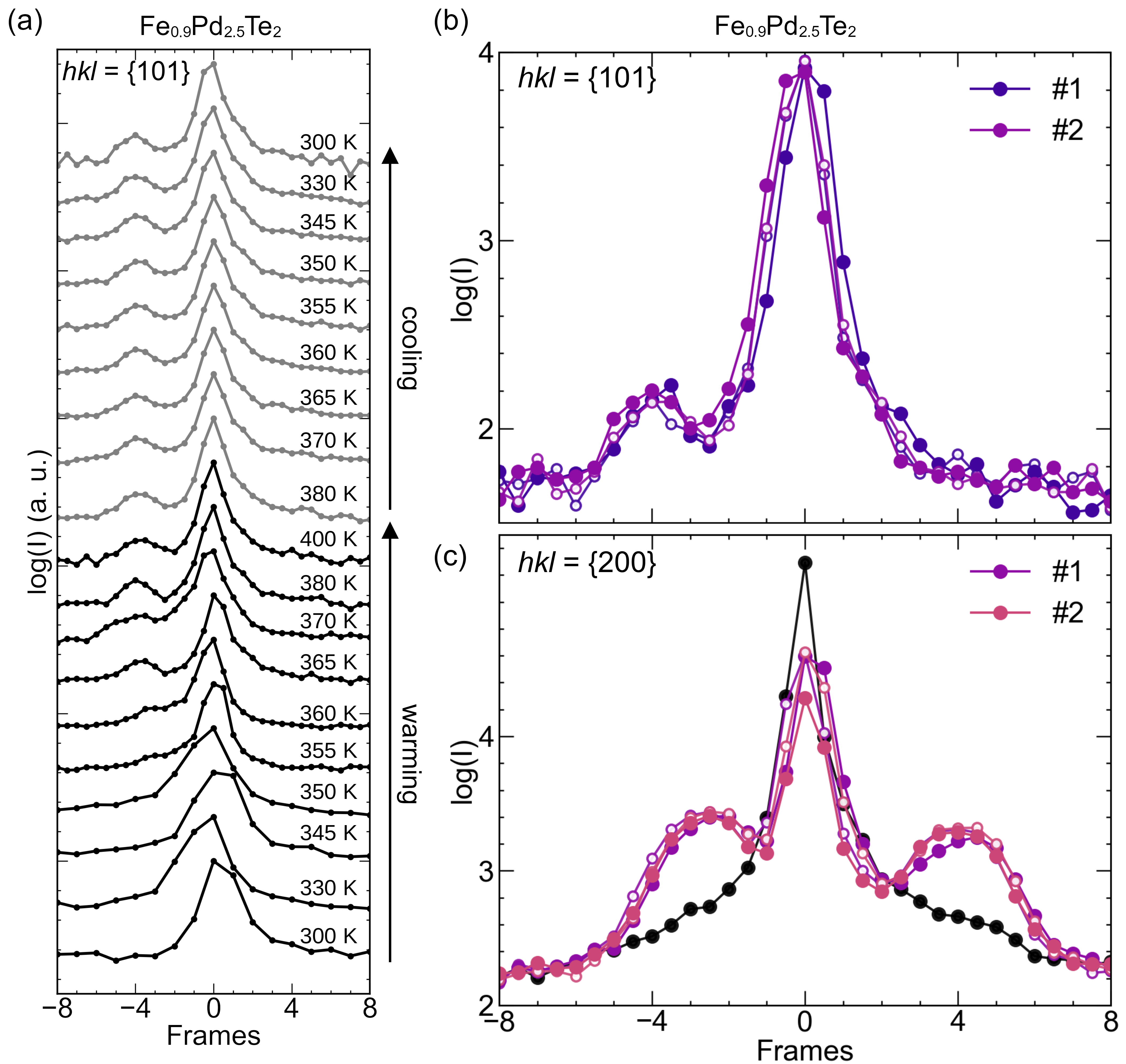}
    \caption{Temperature dependent single crystal diffraction for Fe$_{0.9}$Pd$_{2.5}$Te$_2$. (a) Integrated intensity of the reflection (101) upon warming and subsequent cooling. The first measurement at 300\,K was collected using an as-grown crystal. Integrated intensity of the (b) (101) and (c) (200) reflections at two warming/cooling cycles (\#1 and \#2, respectively) between 360\,K (open circles) and 400\,K (solid circles). The black dots show the (200) integrated intensity measured at 300\,K in an as-grown crystal.}     
    \label{fig:SCXRD_cycles}
\end{figure}

To explore these changes further, Figure \ref{fig:SCXRD_cycles}(a) presents the integrated intensity of the reflection (101) as a function of temperature. The curves are the integration of single crystal data performed in CrysAlis Pro, with frame zero referring to the frame where the intensity of this reflection reached its maximum. Initially, upon measuring at 300 K and prior to heating, we observe only a single peak, within the resolution of our experiment. Above $T\approx370\,$K, this peak splits into two. Increasing the temperature to 400\, K (our experimental limit) only results in a slight increase in the intensity of the secondary peak. Upon cooling, no significant intensity variation was observed, suggesting an irreversible splitting of peaks upon warming. Similar splitting is observed for the \{200\} reflections. After the splitting occurs, subsequent warming/cooling cycles through the transition do not seem to have any effect, as shown in Figs. \ref{fig:SCXRD_cycles}(b-c). We were unable to index these new reflections, suggesting that they do not represent a modulation of the high-temperature structure. 

\begin{figure*}
    \centering
    \includegraphics[scale=.4]{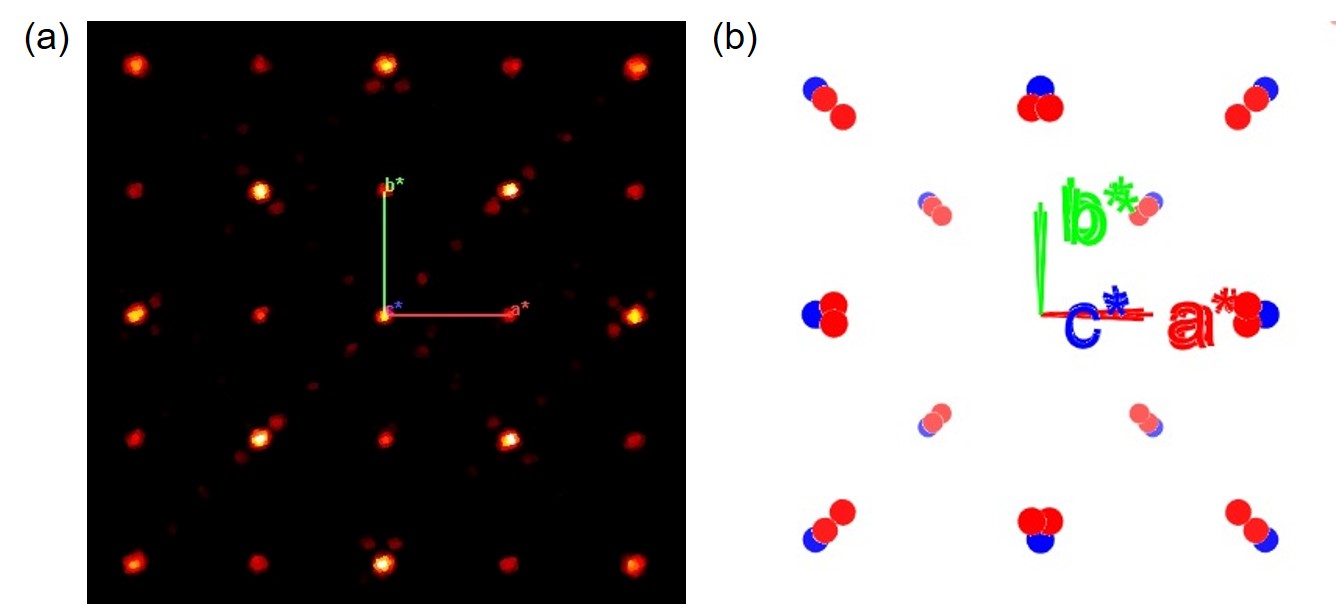}
    \caption{Domain formation in the high-temperature phase of $y>2$ Fe$_x$Pd$_y$Te$_2$ single crystals. (a) Experimental and (b) simulated single crystal diffraction patterns of Fe$_{0.9}$Pd$_{2.5}$Te$_2$ along the 001 direction at 400\,K. Blue dots are the main reflections of the high-temperature tetragonal structure (P$4/nmm$), while red dots show the peak positions associated with two structural domains that exhibit a $\pm$ 5 degree misorientation relative to the tetragonal unit cell ($\gamma = 85^\circ$ and $95^\circ$, respectively).} 
    \label{twinning_400K}
\end{figure*}

We believe that the peak splitting observed for $y >2$ samples above the transition is associated with misaligned domains that form when warming through the first-order structural transition. To support this, Fig. \ref{twinning_400K} compares the experimental data with a simulation of the ($hk0$) plane that superimposes the diffraction patterns expected for the high-temperature tetragonal structure (in blue) with those of two structural domains (in red) that have a $\pm$ 5 degree misorientation relative to the tetragonal structure. Clearly, even this simple picture provides an approximate description of the diffraction pattern observed at 400 K. 

\begin{figure*}[b]
    \centering
    \includegraphics[scale=1.07]{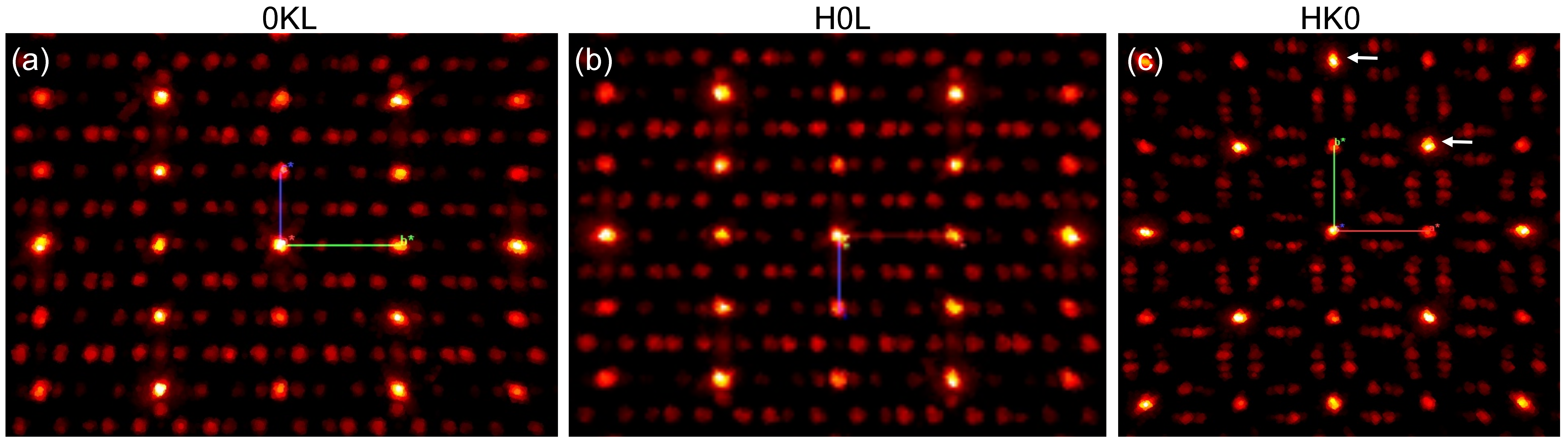}
    \caption{Room temperature single crystal diffraction obtained on a Fe$_{0.9}$Pd$_{2.5}$Te$_2$ sample that was annealed at 400$^\circ$C for $\approx$ 8 h along (a) 100, (b) 010, and (c) 001 directions. White arrows at the HK0 direction indicate the (020) and (110) reflections, where peak splitting is not observed.} 
    \label{fig:SCXRD_SI_annealed}
\end{figure*}

Given the van der Waals nature of Fe$_x$Pd$_y$Te$_2$, it is reasonable to imagine that strains incurred upon rearranging the structure during the first-order transition could result in structural domains associated with a slight misalignment between vdW layers. In support of this hypothesis, Fig. \ref{fig:SCXRD_SI_annealed} shows SCXRD data collected at 300 K on a sample that was annealed at 400 $^\circ$C (673 K) for 8 h. Given that we observe the peak splitting (Figs. \ref{fig:SCXRD_SI} and \ref{fig:SCXRD_cycles}) within minutes of heating above 370 K, and that the split peaks remain when the sample is cooled back to 300 K, we would anticipate that a similar transformation would occur when warmed to 400 $^\circ$C. Yet, data obtained on the annealed sample, shown in Figure \ref{fig:SCXRD_SI_annealed}, does not show any anomalous peak splitting. If we consider the split peaks to be associated with domains of the tetragonal structure, where there is a small rotational misorientation between the vdW layers of the respective domains, high-temperature annealing could likely relax the strains preserving these domains and therefore eliminate the peak splitting.

Lastly, we emphasize that whereas we do not observe the peak splitting described above in the samples that were annealed at 400$^{\circ}$C, Fig. \ref{fig:SCXRD_cycles} shows that, in addition to the structural modulation with $q\approx \big(-0.42\,\,0.17\,\,\frac{1}{2}\big)$, we also observe satellites peaks with $q\approx \big(0.42\,\,0.17\,\,\frac{1}{2}\big)$, that is, with a wave vector rotated by 90$^\circ$ around the $c$-axis. This picture is consistent with the 90$^\circ$ twinning reported for FePd$_2$Te$_2$ \cite{shi2024} and is very likely the result of the symmetry breaking that occurs when cooling through the transition.

\begin{table*}[htbp] 
    \centering
    \begin{ruledtabular}
    \caption{Single crystal data and structural refinement information for Fe$_{0.9}$Pd$_{2.5}$Te$_2$ sample.}
    \begin{tabular}{lll}
                                       & \multicolumn{1}{c}{400 K}      & \multicolumn{1}{c}{300 K}      \\ \hline
Refined formula                             & Fe$_{1.02}$Pd$_{2.32}$Te$_2$  & Fe$_{0.93}$Pd$_{2.55}$Te$_2$  \\
F.W (g/mol)                                 & 560.58                        & 578.2                         \\
Space group, $Z$                            & $P4/nmm$, 1                   & $P2/n(a1/2g)00$, 1            \\
$a = b$ (\AA)                                 & 4.0067 (1)                    & 4.0125 (3)                    \\
$c$ (\AA)                                 & 6.5279 (4)                    & 6.5192 (5)                    \\
$\beta$ ($^\circ$)                           & 90                            & 90.146 (7)                    \\
$\theta$ range ($^\circ$)                    & 2.49-31.287                  & 2.19-32.13                    \\
No. reflections, $R_{int}$                  & 3029, 0.0225                  & 48785, 0.0739                         \\
No. independent reflections                 & 238                           & 672                           \\
No. parameters                              & 11                            & 19                            \\
$R_1$, $wR_2$ [$I>2d(I)$]      & 0.0185, 0.0465                & 0.0158, 0.0529                \\
Goodness of fit                             & 1.162                        & 2.1141                        \\
Diffraction peak and role ($e^{-}/\AA^{3}$) & 1.879, -1.439                 & 0.40, -0.75                  
\end{tabular}
\label{tab:refinementZX679}
    \end{ruledtabular}
\end{table*}

\begin{table*}[htbp]
    \centering
    \begin{ruledtabular}
    \caption{Single crystal data and structural refinement information for FePd$_{2.3}$Te$_2$ sample.}
    \begin{tabular}{lll}
                                            & \multicolumn{1}{c}{400 K}     & \multicolumn{1}{c}{300 K}    \\ \hline
Refined formula                             & Fe$_{1.03}$Pd$_{2.33}$Te$_2$ & Fe$_{1.09}$Pd$_{2.23}$Te$_2$ \\
F.W (g/mol)                                 & 560.13                        & 552.8                        \\
Space group, $Z$                            & $P4/nmm$, 1                   & $P2/n(a1/2g)00$, 1           \\
$a=b$ (\AA)                                 & 4.0092 (2)                    & 4.0012 (5)                   \\
$c$ (\AA)                                 & 6.5284 (6)                    & 6.5210 (6)                   \\
$\beta$ ($^\circ$)                           & 90                            & 90.15(1)                   \\
$\theta$ range ($^\circ$)                    & 2.432-31.338                  & 2.19-49.24                   \\
No. reflections, $R_{int}$                  & 2455, 0.0197                  & 42920, 0.0721                        \\
No. independent reflections                 & 234                           & 683                          \\
No. parameters                              & 11                            & 18                           \\
$R_1$, $wR_2$ [$I>2d(I)$]       & 0.0199, 0.0532                 & 0.029, 0.0857               \\
Goodness of fit                             & 1.244                         & 3.044                       \\
Diffraction peak and role ($e^{-}/\AA^{3}$) & 0.971, -2.073                 & 0.56, -2.17                 
\end{tabular}
\label{tab:refinementZX832}
    \end{ruledtabular}
\end{table*}

\begin{table*}[htbp]
    \centering
    \begin{ruledtabular}
    \caption{Single crystal data and structural refinement information for Fe$_{1.1}$Pd$_{1.8}$Te$_2$ sample.}
    \begin{tabular}{lll}
                                            & \multicolumn{1}{c}{400 K}     & \multicolumn{1}{c}{300 K}    \\ \hline
Refined formula                             & FePd$_{2.04}$Te$_2$           & FePd$_{2.02}$Te$_2$ \\
F.W (g/mol)                                 & 528.64                        & 528.11                        \\
Space group, $Z$                            & P$2_1/m$, 2                   & P$2_1/m$, 2           \\
$a$ (\AA)                                 & 7.5716 (12)                    & 7.5311 (17)                   \\
$b$ (\AA)                                 & 3.9679 (4)                     & 3.9504 (6) \\
$c$ (\AA)                                 & 7.6876 (12)                    & 7.7241 (16)                   \\
$\beta$ ($^\circ$)                           & 117.60 (2)                    & 118.11 (3)                   \\
$\theta$ range ($^\circ$)                    & 2.359-31.704                  & 2.359-31.696                   \\
No. reflections, $R_{int}$                  & 10832, 0.0793                  & 4357, 0.0774                        \\
No. independent reflections                 & 1399                           & 1338                          \\
No. parameters                              & 40                            & 40                           \\
$R_1$, $wR_2$ [$I>2d(I)$]                   & 0.0844, 0.2343                 & 0.1149, 0.3035               \\
Goodness of fit                             & 1.173                         & 1.079                      \\
Diffraction peak and role ($e^{-}/\AA^{3}$) & 5.72, -4.92                 & 21.34, -4.67                 
\end{tabular}
\label{tab:refinementAB065}
    \end{ruledtabular}
\end{table*}



\begin{table}[htbp] 
\begin{ruledtabular}
\caption{\footnotesize{Atomic coordinates, occupancies, and equivalent isotropic displacement parameters (\AA$^2$) of Fe$_x$Pd$_{y}$Te$_2$ single crystals ($x=0.9-1.1$ and $y=2.5-1.8$) at 400\,K. In the high-temperature phase, the $y>2$ compounds adopt a tetragonal structure (space group P$4/nmm$). At 400\,K, the Fe$_{1.1}$Pd$_{1.8}$Te$_2$ composition still has monoclinic P$2_1/m$ symmetry. Occupancy uncertainties are given between parentheses. The refined errors on $z$ coordinates (not shown) are in the order of $10^{-5}$ for the $y>2$ compounds.}}
\label{tab:structures}
\resizebox{\linewidth}{!}{\begin{tabular}{lllllll}
\multicolumn{7}{c}{T = 400\,K}    \\ \hline
\multicolumn{7}{c}{Fe$_{0.9}$Pd$_{2.5}$Te$_2$ (P$4/nmm$, $\beta=90^\circ$)}    \\ \hline
Atoms & Wyckoff & $x$  & $y$  & $z$   & Occ.      & $U_{eq}$ \\
Te1   & 2c      & 0.5  & 0    & 0.286 & 1         & 0.021    \\
Fe1   & 2a      & 0    & 0    & 0     & 0.508 (8) & 0.023    \\
Pd1   & 2a      & 0    & 0    & 0     & 0.492 (8) & 0.023    \\
Pd2   & 2c      & 0.5  & 0    & 0.697 & 0.670 (3) & 0.028    \\ \hline
\multicolumn{7}{c}{FePd$_{2.3}$Te$_2$ (P$4/nmm$, $\beta=90^\circ$)}            \\ \hline
Atoms & Wyckoff & $x$  & $y$  & $z$   & Occ.      & $U_{eq}$ \\
Te1   & 2c      & 0.5  & 0    & 0.286 & 1         & 0.021    \\
Fe1   & 2a      & 0    & 0    & 0     & 0.515 (9) & 0.022    \\
Pd1   & 2a      & 0    & 0    & 0     & 0.485 (9) & 0.022    \\
Pd2   & 2c      & 0.5  & 0    & 0.698 & 0.681 (4) & 0.027    \\ \hline
\multicolumn{7}{c}{Fe$_{1.1}$Pd$_{1.8}$Te$_2$ (P$2_1/m$, $\beta=117.6^\circ$)}    \\ \hline
Atoms & Wyckoff & $x$  & $y$  & $z$   & Occ.      & $U_{eq}$ \\
Te1   & 2e      & 0.23 & 0.75 & 0.01  & 1         & 0.023    \\
Te2   & 2e      & 0.73 & 0.75 & 0.52  & 1         & 0.024    \\
Fe1   & 2e      & 0.12 & 0.75 & 0.62  & 1         & 0.017    \\
Pd1   & 2e      & 0.63 & 0.75 & 0.13  & 1         & 0.031    \\
Pd2   & 2e      & 0.03 & 0.75 & 0.23  & 0.75(1)   & 0.024    \\
Pd3   & 2e      & 0.52 & 0.75 & 0.72  & 0.30(1)   & 0.030   
\end{tabular}}
\end{ruledtabular}
\end{table}

\begin{table}[htbp] 
\begin{ruledtabular}
\caption{\footnotesize{Atomic coordinates, occupancies, and equivalent isotropic displacement parameters (\AA$^2$) of Fe$_x$Pd$_{y}$Te$_2$ single crystals ($x=0.9-1.1$ and $y=2.5-1.8$) at 300\,K. In the low-temperature phase, the $y>2$ compounds adopt a \textit{quasi}-tetragonal structure with monoclinic symmetry (space group P$2/n$, n. 13). At 300\,K, the Fe$_{1.1}$Pd$_{1.8}$Te$_2$ composition has monoclinic P$2_1/m$ symmetry (n. 11). Occupancy uncertainties are given between parentheses. The refined errors on $z$ coordinates (not shown) are in the order of $10^{-5}$ for the $y>2$ compounds.}}
\label{tab:structures300K}
\resizebox{\linewidth}{!}{\begin{tabular}{lllllll}\multicolumn{7}{c}{T = 300\,K}                                                     \\ \hline
\multicolumn{7}{c}{Fe$_{0.9}$Pd$_{2.5}$Te$_2$ (P$2/n$, $\beta\approx 90.1^\circ$)} \\ \hline
Atoms    & Wyckoff    & $x$     & $y$     & $z$       & Occ.         & $U_{eq}$    \\
Te1      & 2f         & 0.5     & 0       & 0.287     & 1            & 0.014       \\
Fe1      & 2e         & 0       & 0       & 0.003     & 0.463 (5)    & 0.017       \\
Pd1      & 2e         & 0       & 0       & 0.003     & 0.537 (5)    & 0.017       \\
Pd2      & 2f         & 0.5     & 0       & 0.700     & 0.738 (2)    & 0.020       \\ \hline
\multicolumn{7}{c}{FePd$_{2.3}$Te$_2$ (P$2/n$, $\beta\approx 90.1^\circ$)}         \\ \hline
Atoms    & Wyckoff    & $x$     & $y$     & $z$       & Occ.         & $U_{eq}$    \\
Te1      & 2f         & 0.5     & 0       & 0.286     & 1            & 0.015       \\
Fe1      & 2e         & 0       & 0       & 0.0001    & 0.543 (7)    & 0.017       \\
Pd1      & 2e         & 0       & 0       & 0.0001    & 0.457 (7)    & 0.017       \\
Pd2      & 2f         & 0.5     & 0       & 0.697     & 0.657 (3)    & 0.018       \\ \hline
\multicolumn{7}{c}{Fe$_{1.1}$Pd$_{1.8}$Te$_2$ (P$2_1/m$, $\beta=118.1^\circ$)}     \\ \hline
Atoms    & Wyckoff    & $x$     & $y$     & $z$       & Occ.         & $U_{eq}$    \\
Te1      & 2e         & 0.23    & 0.75    & 0.01      & 1            & 0.017       \\
Te2      & 2e         & 0.73    & 0.75    & 0.52      & 1            & 0.018       \\
Fe1      & 2e         & 0.12    & 0.75    & 0.62      & 1            & 0.014       \\
Pd1      & 2e         & 0.63    & 0.75    & 0.14      & 1            & 0.023       \\
Pd2      & 2e         & 0.04    & 0.75    & 0.23      & 0.85 (1)     & 0.018       \\
Pd3      & 2e         & 0.53    & 0.75    & 0.72      & 0.17 (1)     & 0.019   
\end{tabular}}
\end{ruledtabular}
\end{table}

\clearpage

\section{Curie Weiss fits}

\begin{figure*}[htbp]
    \centering
    \includegraphics{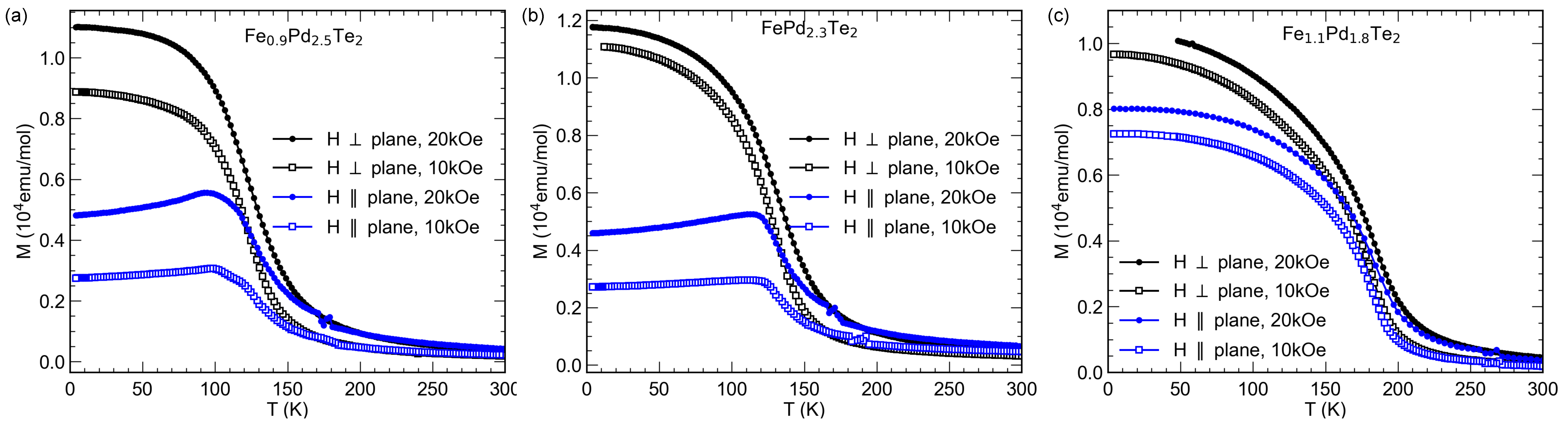}
    \caption{Temperature dependence of molar magnetization $M$ for (a) Fe$_{0.9}$Pd$_{2.5}$Te$_2$, (b) FePd$_{2.3}$Te$_2$, and (c) Fe$_{1.1}$Pd$_{1.8}$Te$_2$ single crystals. All measurements were field-cooled with $H=20\,$kOe (solid circles) or $H=10\,$kOe (open squares). In-plane and out-of-plane measurements are shown in black and blue, respectively. The in-plane orientation is $H\perp c$ for $y>2$ samples and $H\perp [10\bar{1}]$ for Fe$_{1.1}$Pd$_{1.8}$Te$_2$, respectively. Data oscillations around 160\,K and 260\,K in the out-of-plane measurements arise due to zero-crossing artifacts from background subtraction.} 
    \label{fig:figure5_SI}
\end{figure*}

\begin{figure*}[htbp]
    \centering
    \includegraphics{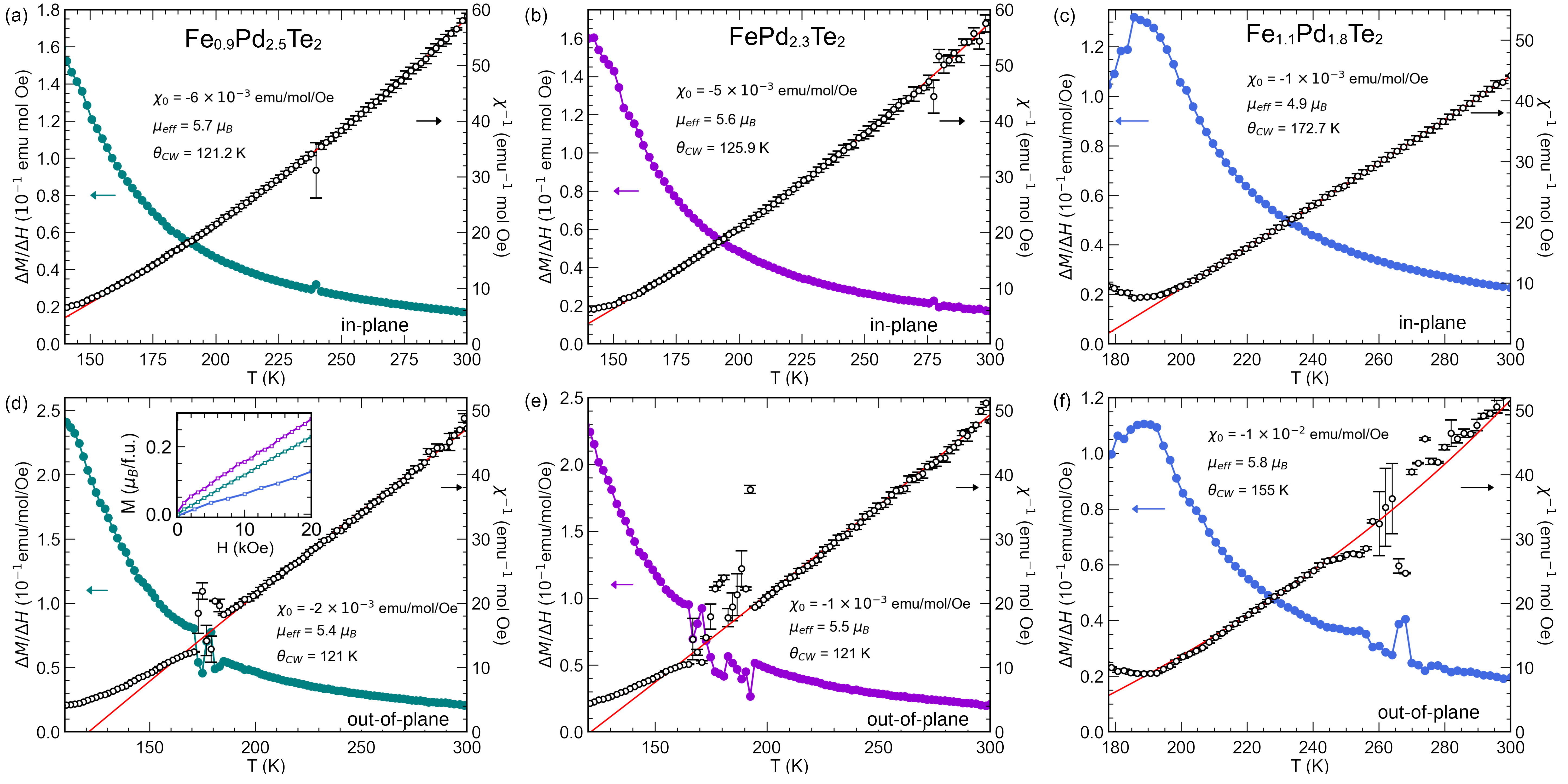}
    \caption{Temperature dependence of magnetic susceptibility $\chi = \Delta M/\Delta H$ (left axis, solid circles) for (a, d) Fe$_{0.9}$Pd$_{2.5}$Te$_2$, (b, e) FePd$_{2.3}$Te$_2$, and (c, f) Fe$_{1.1}$Pd$_{1.8}$Te$_2$ single crystals. The inverse susceptibility (right axis, open circles) and corresponding CW fits (red lines) are shown. All samples show CW behavior above the respective magnetic transition temperatures, either (a-c) in-plane or (d-e) out-of-plane. The in-plane orientation is $H\perp c$ for $y>2$ samples and $H\perp [10\bar{1}]$ for Fe$_{1.1}$Pd$_{1.8}$Te$_2$, respectively. The noise in the data around 160\,K and 260\,K in the out-of-plane measurements are due to zero-crossing artifacts of the raw data from background subtraction due to the use of a diamagnetic disc. The inset on (d) shows a close up view of in-plane field-dependent magnetization isotherms collected at T = 180\,K (for Fe$_{0.9}$Pd$_{2.5}$Te$_2$ and FePd$_{2.3}$Te$_2$) or T = 250\, K (for Fe$_{1.1}$Pd$_{1.8}$Te$_2$).} 
    \label{fig:figure4_SI}
\end{figure*}

To determine the effective moments ($\mu_{eff}$) in the paramagnetic state, we fit the magnetization data to the Curie-Weiss (CW) law,

\begin{center}
\begin{equation*}
\chi = \frac{C}{T-\Theta} + \chi_0
\end{equation*}
\end{center}

\noindent where $C$ is the Curie constant, $\Theta$ is the Weiss temperature, and $\chi_0$ is the temperature independent contribution to the susceptibility. As seen in the inset of Fig. \ref{fig:figure4_SI}(d), the magnetization above \textit{T}$_C$ is not linear at small fields ($H<\,$5\,kOe), likely due to a small amount of ferromagnetic impurity, so the expression $\chi = M/H$ is not valid. Therefore, we measured temperature dependent magnetization at \textit{H} = 10 kOe and 20 kOe, where the data is shown in Figure \ref{fig:figure5_SI}(a-c). As these fields are in the linear \textit{M}(\textit{H}) regime in the paramagnetic state, so we estimated the magnetic susceptibility as $\chi\approx\Delta M/\Delta H$, where $\Delta M = M(\text{20 kOe}) - M(\text{10 kOe})$ and $\Delta H =\text{20 kOe} - \text{10 kOe}$. The \textit{M}(\textit{T}) data shown in Figure 3 of the main text was measured at \textit{H} = 10 kOe, corresponding to the open symbol curves in Fig. \ref{fig:figure5_SI}.

Figure \ref{fig:figure4_SI} shows the dependence of $\chi$ (left axis, colored data) for Fe$_x$Pd$_{y}$Te$_2$ single crystals. By fitting the $\chi^{-1}$ curves (right axis, open circles) to the CW law over 200-300\,K, we estimated the $\mu_{eff}$ and $\Theta$ values that were presented in Table II of the main text.

\newpage

\bibliography{bibSI}